\documentclass[12pt]{iopart}
\usepackage{iopams}
\usepackage{setstack}
\usepackage[colorlinks=true,citecolor=blue,linkcolor=magenta]{hyperref}
\usepackage{graphicx}
\usepackage{enumitem} 
\usepackage{dcolumn}
\usepackage{bm}
\usepackage{dsfont}
\usepackage{color}
\def\ii{{\rm i}}

\def\kg{k_{\rm 1D}} 
\def\braket#1{\mathinner{\langle{#1}\rangle}}

\newcommand{\ket}[1]{|#1\rangle}
\newcommand{\ex}[1]{\langle #1 \rangle}

\newcommand{\tbl}[1]{#1}
\newcommand{\tblue}[1]{#1}

\usepackage{cite}
\begin{document}
\title{Subradiant states of quantum bits coupled to a one-dimensional waveguide}
\author{Andreas Albrecht$^{1*}$, Lo\"{i}c Henriet$^{1*}$, Ana Asenjo-Garcia$^{1,2,3\ddagger}$, Paul B. Dieterle$^{3,4\dagger}$, Oskar Painter$^{3,4}$, and Darrick E. Chang$^{1,5}$ }
\address{$^1$ ICFO-Institut de Ciencies Fotoniques, The Barcelona Institute of Science and Technology, 08860 Castelldefels (Barcelona), Spain}
\address{$^2$ Norman Bridge Laboratory of Physics MC12-33, California Institute of Technology, Pasadena, CA 91125, USA}
\address{$^3$ Institute for Quantum Information and Matter, California Institute of Technology, Pasadena, CA 91125, USA}
\address{$^4$ Kavli Nanoscience Institute and Thomas J. Watson, Sr., Laboratory of Applied Physics, California Institute of Technology, Pasadena, CA 91125, USA}
\address{$^5$ ICREA-Instituci\'{o} Catalana de Recerca i Estudis Avan\c{c}ats, 08015 Barcelona, Spain}
\address{$^*$ A. Albrecht and L. Henriet contributed equally to this work}
\address{$^\dagger$ Current address: Department of Physics, Harvard University, Cambridge, MA 02138, USA}
\address{$^\ddagger$ Current address: Department of Physics, Columbia University, New York, NY 10027, USA}

\begin{abstract}

The properties of coupled emitters can differ dramatically from those of their individual constituents. Canonical examples include sub- and super-radiance, wherein the decay rate of a collective excitation is reduced or enhanced due to correlated interactions with the environment. Here, we systematically study the properties of collective excitations for regularly spaced arrays of quantum emitters coupled to a one-dimensional (1D) waveguide. We find that, for low excitation numbers, the modal properties are well-characterized by spin waves with a definite wavevector. Moreover, the decay rate of the most subradiant modes obeys a universal scaling with a cubic suppression in the number of emitters. Multi-excitation subradiant eigenstates can be built from fermionic combinations of single excitation eigenstates; such ``fermionization'' results in multiple excitations that spatially repel one another. We put forward a method to efficiently create and measure such subradiant states, which can be realized with superconducting qubits. These measurement protocols probe both real-space correlations (using on-site dispersive readout) and temporal correlations in the emitted field (using photon correlation techniques).

\end{abstract}

\maketitle

\section{Introduction}
Superconducting qubits coupled to photons propagating in open transmission lines \cite{astafiev10,wQED_circuit_2,van_Loo_Science} offer a platform to realize and investigate the fascinating world of quantum light-matter interactions in one dimension -- so-called ``waveguide quantum electrodynamics~(QED)''\,\cite{chang07,Shen07,vetsch04,zheng10,pichlerPRA15,sanchez-burillo17,Roy17,GU17}. Such systems enable a number of exotic phenomena that are difficult to observe or have no obvious analogue in other settings, such as near-perfect reflection of light from a single resonant qubit \cite{shen05,chang07,astafiev10,abdumalikov10,wQED_circuit_2,Hoi13}, or the dynamical Casimir effect\,\cite{Wilson2011}, \tbl{and allow for the measurement of quantum vacuum fluctuations\,\cite{Hoi15}}. One particularly interesting feature of these systems is that the interaction between multiple qubits, mediated by photon absorption and re-emission, is of infinite range. This can give rise to strong collective effects in multi-qubit systems\,\cite{tudela13,pichlerPRA15,Fang15}. For example, it has been observed that two qubits separated by a substantial distance can exhibit super- or sub-radiance, wherein a single collective excitation can decay at a rate faster or slower than that of a single qubit alone \cite{van_Loo_Science}.

The physics associated with collective effects in waveguide QED has attracted growing interest, and there have been a number of proposals that implicitly exploit sub- and super-radiant emission to realize atomic mirrors\,\cite{chang12}, photon Fock state synthesis\,\cite{Tudela17}, or quantum computation\,\cite{dzsotjan2010,Paulisch16}. The fundamental properties of the qubit modes themselves, such as their spatial character and decay spectrum, have been studied recently in the classical single-excitation regime \cite{Haakh2016}.

Here, we aim to provide a systematic description of single- and multi-excitation subradiant states in ordered arrays by using a spin-model formalism, wherein emission and re-absorption of photons by qubits is exactly accounted for. Our study reveals a number of interesting characteristics. In particular, as the number of qubits $N$ increases, we show that the Liouvillian ``gap'' closes, i.e., there exists a smooth distribution of decay rates associated with subradiant states whose value approaches zero. Furthermore, we find that the most subradiant multi-excitation states exhibit ``fermionic'' correlations in that the excitations obey an effective Pauli exclusion principle. \tbl{These calculations parallel a similar investigation involving subradiant states of an ordered chain of atoms in three-dimensional space\,\cite{asenjo17}. The finding of similar properties suggests a certain degree of ``universality'' to the phenomenon of subradiance.} Next, we propose a realistic experimental protocol to measure these exotic spatial properties, and finally investigate the correlations in the corresponding emitted field. Taken together, these results show that the physics of subradiance is itself a rich many-body problem.

\begin{figure}[htb]
\begin{centering}
\includegraphics[scale=0.35]{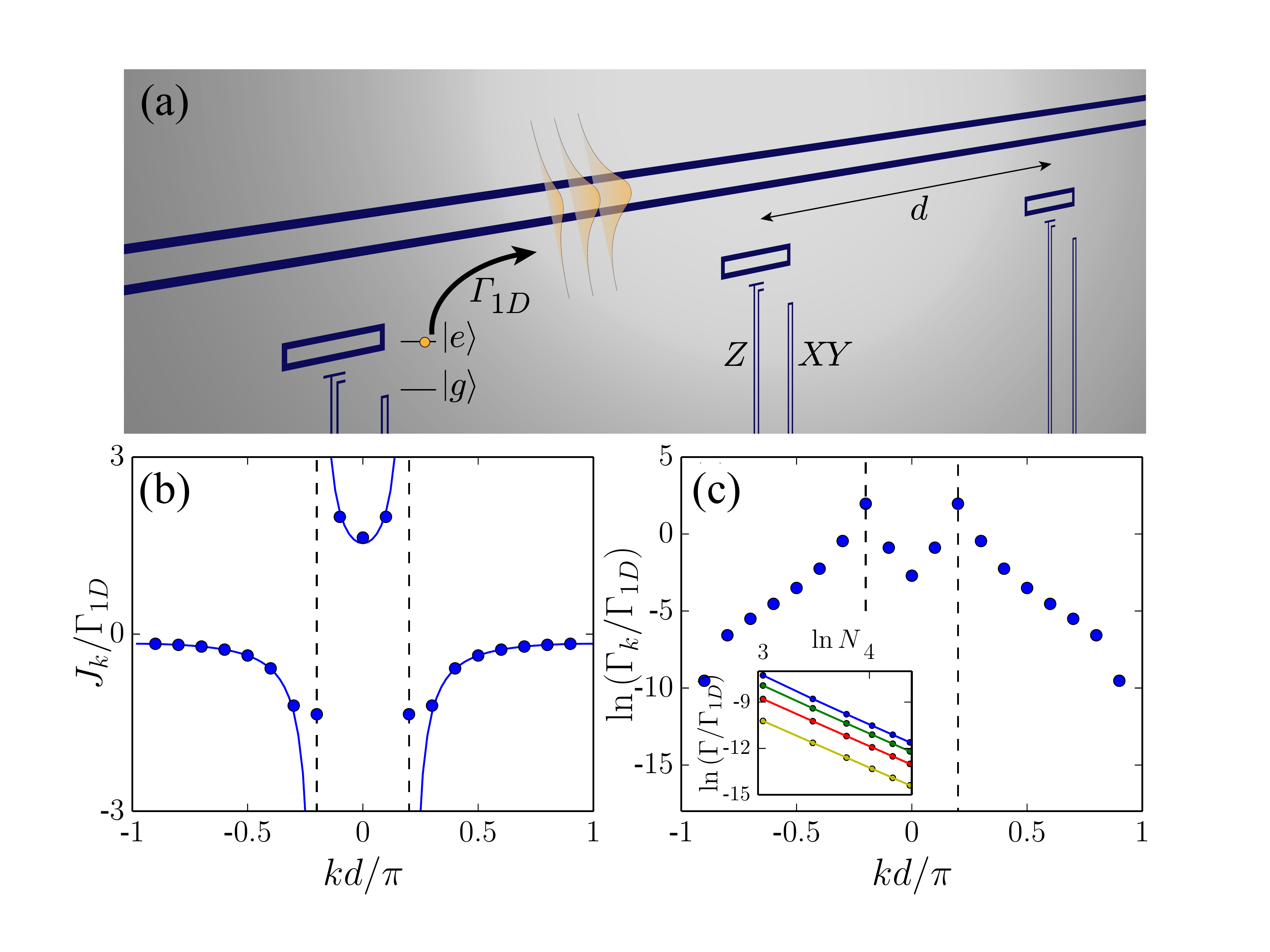}
\caption{\label{Fig_1}\textbf{(a)} Schematic of planar transmon qubits capacitively coupled to a coplanar waveguide. Photon-mediated interactions couple the qubits together with an amplitude determined by the single-qubit emission rate $\Gamma_{\rm 1D}$ into the waveguide, and a phase determined by the phase velocity of the transmission line and the distance between qubits.  Collective frequency shifts \textbf{(b)} and decay rates \textbf{(c)} for qubits coupled through a waveguide with $\kg d/\pi=0.2$. Blue circles correspond to the results for a finite system with $N$ = 30 qubits. Black dashed lines correspond to $k=\pm \kg $. The frequency shift for the infinite chain is denoted by the solid line. The inset in \textbf{(c)} shows the scaling of the decay rate with qubit number $\Gamma/\Gamma_{\rm 1D}\sim N^{-3}$ for the 4 most subradiant states.}
\end{centering}
\end{figure}

\section{Eigenmodes of the atom-waveguide system and collective emission properties}
\subsection{Setup and spin model description}
 We consider $N$ regularly-spaced two-level transmon qubits\,\cite{koch07} with ground and excited states $|g\rangle$, $|e\rangle$ and resonance frequency $\omega_{eg}$. The qubits are dipole coupled to an open transmission line supporting a continuum of left- and right-propagating modes with linear dispersion and velocity $v$ [see Fig.\,\ref{Fig_1}\,(a)]. Integrating out the quantum electromagnetic environment in the Markovian regime, one finds that emission of photons into the waveguide leads to cooperative emission and exchange-type interactions between the qubits \cite{dzsotjan2010,chang12,Lalumiere_PRA,Tommaso_NJP_2015,asenjo17}. The dynamics of the qubit density matrix $\rho$ can be described by a master equation of the form 
\begin{equation}\label{master_eqn} \dot{\rho}=-(i/\hbar)[\mathcal{H}_{\rm eff}\rho-\rho \mathcal{H}_{\rm eff}^{\dagger}]+\sum_{m,n}\Gamma_{m,n}\,\sigma^m_{ge}\rho \sigma^n_{eg}\,,\end{equation}
where the effective (non-Hermitian) Hamiltonian reads\,\cite{chang12,Lalumiere_PRA,Tommaso_NJP_2015}
\begin{equation}
\mathcal{H}_{\rm eff}=\hbar \sum_{m,n=1}^N \left( J_{m,n}-i\frac{\Gamma_{m,n}}{2} \right) \sigma^n_{eg} \sigma^m_{ge}\label{eq:Effective_Hamiltonian},
\end{equation}
with $J_{m,n}=\Gamma_{\rm 1D}  \sin(\kg |z_{m}-z_{n}|)/2$ and $\Gamma_{m,n}= \Gamma_{\rm 1D}  \cos(\kg |z_{m}-z_{n}|)$ denoting the coherent and dissipative interaction rates, respectively. Here, $\Gamma_{\rm 1D}$ is the single qubit emission rate into the transmission line, $\kg=\omega_{eg}/v$ is the resonant wavevector, $\sigma_{\alpha \beta}^m=|\alpha_m \rangle \langle \beta_m|$ acts on the internal states $\{\alpha,\beta\}\in \{g,e\}$ of qubit $m$ at position $z_m=md$, with $d$ the inter-qubit distance. The photonic degrees of freedom can be recovered after solving the qubit dynamics \cite{chang12,Tommaso_NJP_2015}. In particular, the positive-frequency component of the left- and right-going field emitted by the qubits reads
\begin{equation} \hat{E}_{L/R}^+(t)=\hat{E}^{\rm in}_{L/R}(z\pm vt)+\ii \frac{\Gamma_{\rm 1D}}{2}\sum_{n=1}^N e^{\ii \kg |z_{L/R}-z_n|}\hat{\sigma}^n_{ge}(t)\,, \end{equation}
where the field $\hat{E}_{L}^+$ ($\hat{E}_{R}^+$) is measured directly beyond the first (last) qubit, at position $z_L=d$ ($z_R=N d$). Here, $\hat{E}^{\rm in}_{L/R}$ denotes the quantized input field. 

The Markov approximation holds when retardation effects are negligible, that is, when the timescale $L/v$ for a photon to travel within the qubit chain of length $L$ is small as compared to the timescale $\Gamma_{\rm 1D}^{-1}$ of qubit-photon interactions. This condition amounts to $L\ll 10$ m, for typical values of $v\simeq 10^8$ m.s$^{-1}$ and $\Gamma_{\rm 1D}\simeq 10^7$ Hz.

For a given number of excitations, the effective Hamiltonian $\mathcal{H}_{\rm eff}$ defines a complex symmetric matrix that can be diagonalized to find collective qubit modes with complex eigenvalues defining their resonance frequencies (relative to $\omega_{eg}$) and decay rates.

\subsection{The Dicke limit}
In the simple case of $k_{\rm 1D}\,d=2n\pi$ [$(2n+1)\pi$], with $n$ an integer, the coherent qubit-qubit interactions $J_{m,n}$ vanish and the effective Hamiltonian is purely dissipative, \begin{equation} \mathcal{H}_{\rm eff}=-i\frac{\hbar N\Gamma_{\rm 1D}}{2}\, S_{k=0 [k=\pi/d]}^{\dagger}S_{k=0 [k=\pi/d]}\,,\end{equation}
where $S_{k}^\dagger=1/\sqrt{N}\sum_n e^{ik z_n}\sigma_{\rm eg}^n$. The $k=0~ [k=\pi/d]$ collective mode emits superradiantly to the waveguide at a rate $N\Gamma_{\rm 1D}$, while all other modes are dark, with decay rate $\Gamma=0$. This realizes the ideal Dicke model of superradiance\,\cite{dicke54}. Within the setting of a 1D waveguide, it has also been shown that this configuration has interesting quantum optical functionality. For example, the qubits act as a nearly perfect mirror for near-resonant photons\,\cite{chang12, corzo16, sorensen16} and can generate arbitrary photon Fock states on demand\,\cite{gonzalez_tudela2015}.

Away from this spacing, the system becomes multimode\,\cite{Haakh2016}, and results in interesting properties for the single- and multi-excitation eigenstates.

\subsection{Single-excitation modes} 
Numerical diagonalization of $\mathcal{H}_{\rm eff}$ in the single-excitation sector gives $N$ distinct eigenstates $|\psi^{(1)}_{\xi}\rangle=S^{\dagger}_{\xi} |g\rangle^{\otimes N}=\sum_{n} c^{\xi}_n |e_n\rangle$ that obey 
\begin{equation}\mathcal{H}_{\rm eff}\,|\psi^{(1)}_{\xi}\rangle=\hbar\,(J_{\xi}-i\Gamma_{\xi}/2)\,|\psi^{(1)}_{\xi}\rangle.\,\end{equation}
\tbl{Here, $|e_n\rangle=\sigma_{\rm eg}^n|g\rangle^{\otimes N}$ corresponds to having atom $n$ excited, and $J_{\xi}$ and $\Gamma_{\xi}$ represent the frequency shift and decay rate associated with $|\psi^{(1)}_{\xi}\rangle$. Their interpretation as shifts and decay rates can be understood from the equivalent quantum jump interpretation \,\cite{Daley14} of the master equation (\ref{master_eqn}). In particular, within the jump formalism, a wave function evolves under the Schr\"{o}dinger equation governed by $\mathcal{H}_{\rm eff}$, and thus, an eigenstate $|\psi^{(1)}_{\xi}\rangle$ evolves in time as $\exp\left[(-iJ_{\xi}-\Gamma_{\xi}/2)t\right]|\psi^{(1)}_{\xi}\rangle$. The loss of amplitude at a rate $\Gamma_{\xi}$ during evolution is supplemented by quantum jump operators stochastically applied to the wave function (corresponding to the last term 
$\sum_{m,n}\Gamma_{m,n}\,\sigma^m_{ge}\rho \sigma^n_{eg}$ in Eq.\,(\ref{master_eqn})), which physically describes the new state following the decay of an excitation.}

For our particular system of interest, we obtain a broad distribution of decay rates defining superradiant ($\Gamma_{\xi}>\Gamma_{\rm 1D}$) and subradiant ($\Gamma_{\xi}<\Gamma_{\rm 1D}$) states. Ordering the eigenstates by increasing decay rates, i.e., from $\xi=1$ for the most subradiant to $\xi=N$ for the most radiant, we find that strongly subradiant modes exhibit a decay rate $\Gamma_{\xi} \ll \Gamma_{\rm 1D}$ that is suppressed with qubit number as $\Gamma_{\xi}/\Gamma_{\rm 1D}\propto \xi^2/N^3$\,\cite{Tsoi08,Fang15}. \tbl{This decay rate scaling is similar to the case of a 1D chain of atoms in  \tblue{3D} free space with lattice spacing smaller than half of the transition wavelength\,\cite{asenjo17}, while in the present case there is no restriction on the lattice constant other than not being in the Dicke limit discussed earlier}. \tblue{Such a cubic scaling is rather generic to so-called 1D ``boundary dissipation'' models \,\cite{Hafezi12,Znidaric15,asenjo17}, where losses occur solely at the ends of the physical system. In our system, the periodic chain of qubits guides light perfectly in the form of polaritons, which are then dissipated into the waveguide when they hit the ends of the chain.}

An interesting consequence of \tblue{the scaling of $\Gamma_{\xi}$ with $N$ for the most subradiant states} is that, in the thermodynamic limit, the spectrum of decay rates becomes smooth and the ``gap'' of minimum decay rate closes. For an infinite chain, the eigenstates of $\mathcal{H}_{\rm eff}$ take the form of Bloch spin waves $|\psi^{(1)}_k\rangle=S^{\dagger}_{k} |g\rangle^{\otimes N}$, with $k$ being a quantized wavevector within the first Brillouin zone ($|k|\leq \pi/d$). For finite chains, the eigenstates are instead described in momentum space by a wavepacket with a narrow distribution of wavevectors around a dominant wavevector $k$, which can thus serve as an unambiguous label of states. In Fig.\,\ref{Fig_1}\,(b) and (c), we show the distribution of frequency shifts $J_k$ and decay rates $\Gamma_k$ with $k$ for $N$=30 qubits and $\kg d/\pi=0.2$. We find large decay rates and frequency shifts for eigenstates with wavevectors $k$ close to the resonant wavevectors $\pm k_{\rm 1D}$. Conversely, we obtain decay rate minima and small frequency shifts around $kd=0$ and $|k|d=\pi$. For $k_{\rm 1D}\,d>0.5\pi$ ($k_{\rm1D}\,d<0.5\pi$), wavevectors $k\,d=0$ form the global (local) and $|k|\,d=\pi$ the local (global) decay rate minimum, respectively.  \tbl{Such a behavior differs from what is found in a free-space atomic chain\,\cite{asenjo17}, where subradiant states are located in a region $|k|>\omega_{eg}/c$. }

The $k$-dependence can be understood by considering the infinite lattice limit, where the qubits and waveguide generally hybridize to form two lossless polariton bands. For an infinite system, the total Hamiltonian describing both the qubits and photonic degrees of freedom is given by
\begin{equation}
\mathcal{H}_{tot}=\sum_{k} \left\{ \hbar\omega_{eg} S^{\dagger}_{k} S_{k} + \hbar \omega_k a_{k}^{\dagger}a_{k}+ \hbar g_k \left[a_{k}S_k^\dagger +h.c.\right]\right\}\label{eq:infinite_main}.
\end{equation}
Here, $S_k^{\dagger}$ creates a collective spin excitation with $k$ a quantized wavevector in the first Brillouin zone, and $a_{k}^{\dagger}$ is the creation operator of a propagating excitation with wave-vector $k$ and frequency $\omega_k=v|k|$ in the transmission line. The third term of Eq.\,(\ref{eq:infinite_main}) describes the interaction between the qubits and the electromagnetic field, where the parameter $g_k$ quantifies the strength of the interaction. We take a light-matter coupling of the form\,\cite{Lalumiere_PRA} $\sum_k g_k^2\, \delta(\omega-\omega_k)=g^2 \omega\, \theta(\omega_f-\omega)$, where $\omega_f>\omega_{eg}$ is a high-frequency cutoff and $\theta(.)$ is the Heaviside step function.

\begin{figure}[h!]
\center
\includegraphics[scale=0.4]{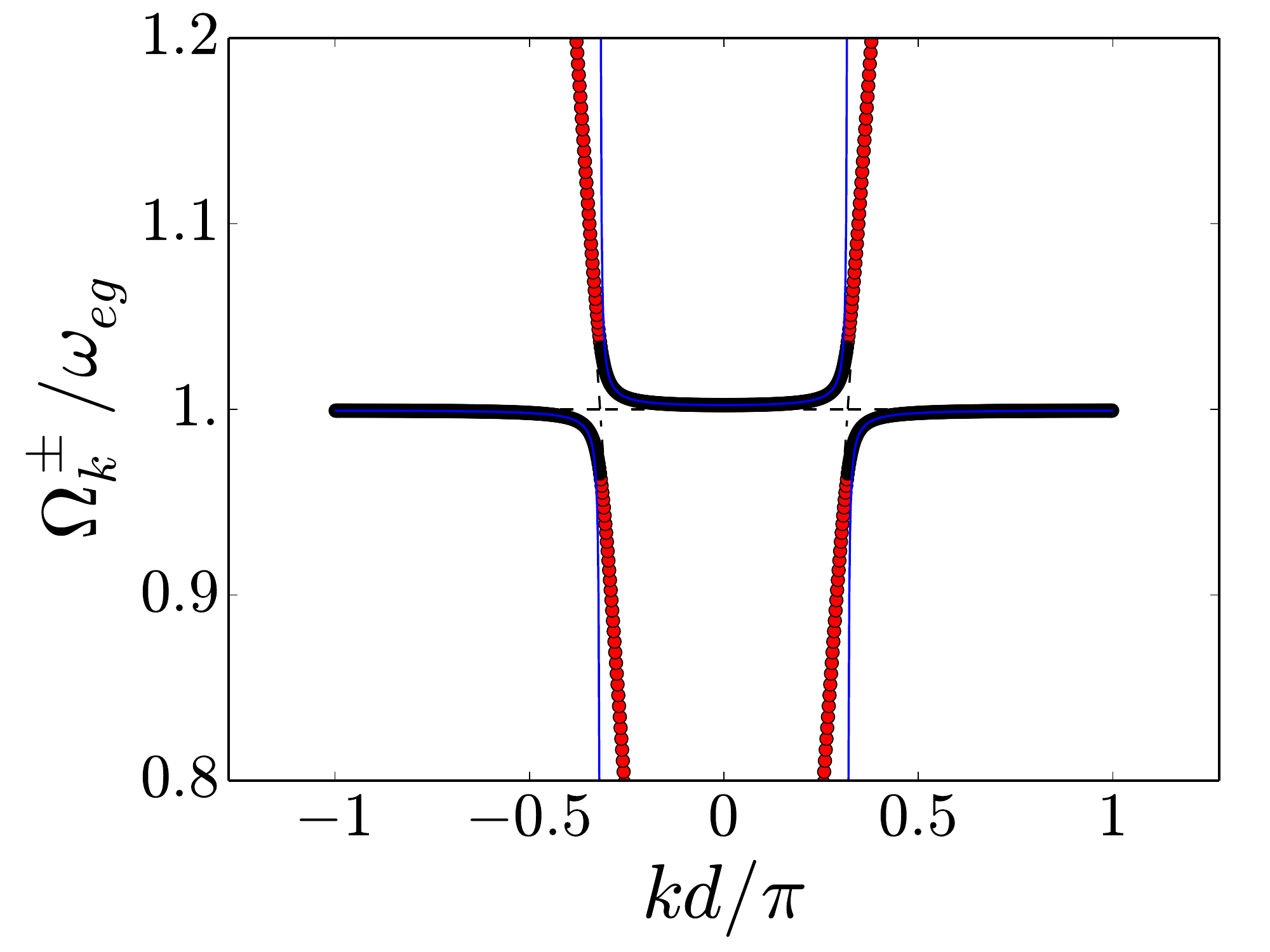}
\caption{The dots show the two eigenvalue solutions of Eq.\,(\ref{eq:infinite_main}), which are plotted in black (red) when the qubit (photon) component of the polariton is the largest in absolute value. The solid blue line corresponds to the result obtained from the direct Bloch diagonalization of $\mathcal{H}_{\rm eff}$, and the dashed black lines show the bare dispersion relations of the isolated qubits and photons. Here, $k_{\rm 1D}d/\pi=0.32$ and $g=0.01$.}
\label{Fig_polariton_main}
\end{figure}

For each wavevector, and within the single-excitation sector, the Hamiltonian\,(\ref{eq:infinite_main}) represents a 2x2 matrix that can be diagonalized to yield frequencies $\Omega^{\pm}_k$, as shown in Fig.\,\ref{Fig_polariton_main}. Physically, the two distinct solutions correspond to a qubit branch and a waveguide branch, with significant hybridization of the two around their intersection at $k=\pm \kg$. For a finite system, this implies that a collective excitation of qubits with wavevector close to $\pm k_{\rm 1D}$ efficiently radiates into the waveguide, as confirmed in Fig.\,\ref{Fig_1}\,(c). Polaritons with wavevector around $k\sim k_{\rm 1D}$ ($k=0,\pi/d$) are most (least) impedance-matched at their boundaries to the dispersion relation of propagating photons in the bare waveguide, thus giving rise to super-radiant (sub-radiant) emission.

In the regions where $|J^{\pm}_k|/\omega_{eg} \ll 1$, with $J^{\pm}_k=\Omega^{\pm}_k-\omega_{eg}$ the frequency shift, we recover a good agreement with the expression obtained from the direct Bloch diagonalization of the effective spin-model Hamiltonian\,(\ref{eq:Effective_Hamiltonian}), which predicts $J_k\sim \Gamma_{\rm 1D}[\cot \bigl((k+\kg)d/2\bigr)+\cot \bigl((\kg-k)d/2\bigr)]/4$ for $k\neq \pm \kg$ and $\Gamma_k \sim N\Gamma_{\rm 1D} \delta_{k,\pm \kg}/2$, with $\Gamma_{\rm 1D} =2 \pi g^2 \omega_{eg}$ (see  Fig.\,\ref{Fig_polariton_main}). That dispersion relation is plotted in Fig.\,\ref{Fig_1}\,(b) as a solid line, and matches well with the frequency shifts obtained for a finite system.  While the single-excitation limit is readily solvable either within the spin model or the full qubit-field Hamiltonian of Eq.\,(\ref{eq:infinite_main}), the spin model is a powerful simplifying tool to understand the properties of multiple excited qubits interacting via common photonic modes.

\subsection{Multi-excitation modes}

A quadratic bosonic Hamiltonian would enable us to easily find the multi-excitation eigenstates of $\mathcal{H}_{\rm eff}$ from the single-excitation sector results. Here, however, the spin nature prevents multiple excitations of the same qubit. Specifically, two-excitation states $|\varphi^{(2)}_{\xi}\rangle=\mathcal{N}_2(S^{\dagger}_{\xi})^2 |g\rangle^{\otimes N}$, with $\mathcal{N}_2$ a normalization factor, are not eigenstates of the effective Hamiltonian\,(\ref{eq:Effective_Hamiltonian}). Moreover, for an index $\xi$ corresponding to a subradiant single-excitation mode, the initial decay rate of $|\varphi^{(2)}_{\xi}\rangle$ is significantly greater than twice the single-excitation decay rate $\Gamma_{\xi}$. 

This discrepancy can be explained by noting that the spatial profile of $|\varphi^{(2)}_{\xi}\rangle$, i.e. the probability $p_{m,n}=|\ex{e_m,e_n|\varphi_{\xi}^{(2)}}|^2$ for qubits $m$ and $n$ to be excited, contains a sharp cut along the diagonal $m=n$ ($p_{m,m}\equiv 0$). In reciprocal space, this corresponds to a broad distribution of wavevector components, including radiant contributions responsible for an increased decay rate. From this qualitative discussion, one expects the excitations forming a multi-excitation subradiant eigenstate to be smoothly repelled from one another.

\begin{figure}[tbh]
\begin{centering}
 \includegraphics[scale=0.5]{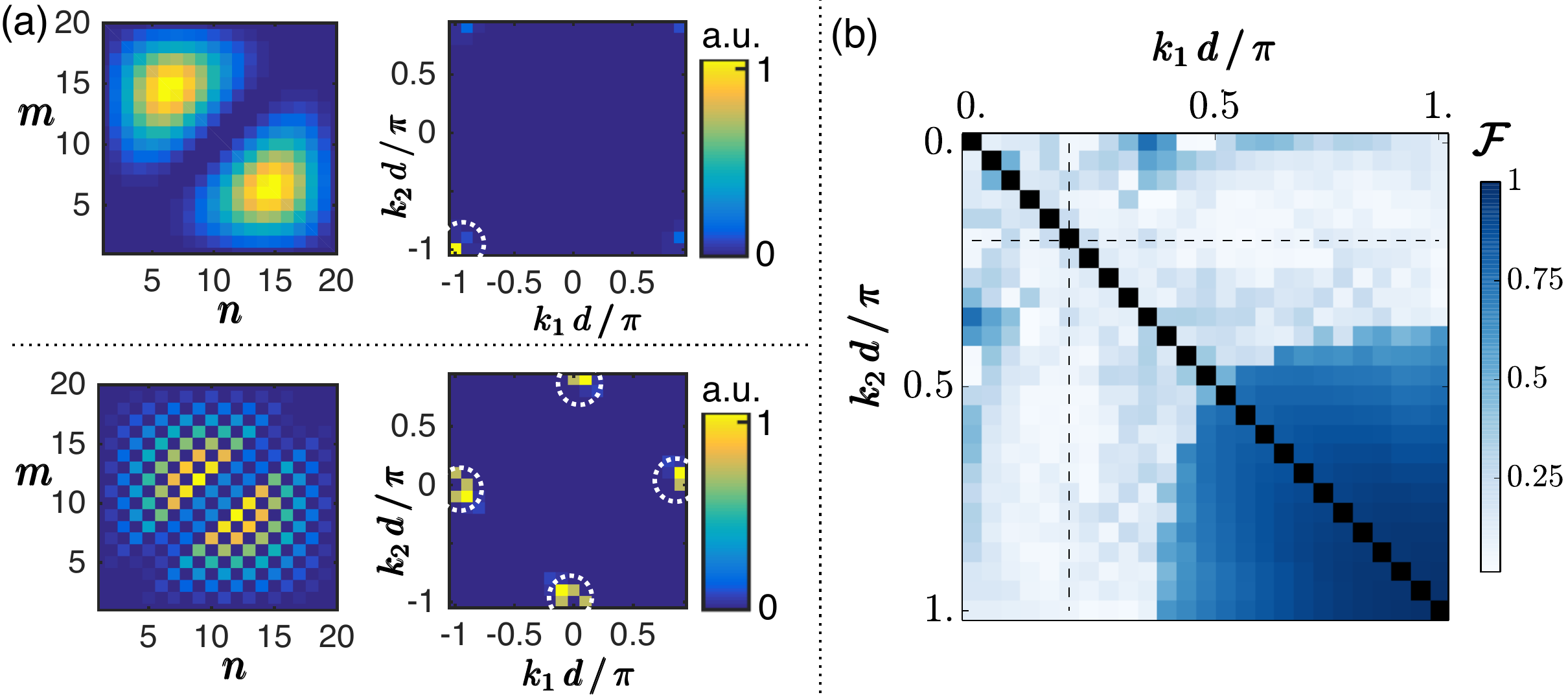}
\caption{\textbf{(a)} Probability amplitude $|c_{mn}|^2$ in real space (left) and in reciprocal space $|c_{k_1,k_2}|^2$ (right) of the wavefunction profile of the most subradiant two-excitation eigenstate for $\kg  d/\pi=0.2$ (top) and $\kg  d/\pi=0.5$ (bottom), for $N=20$ qubits. The amplitude $c_{mn}$ for atoms $m,n$ to be excited is fully specified by $m<n$, but for visual appeal here we symmetrize the plot by taking $c_{mn}=c_{nm}$.  Dotted dashed circles are a guide to the eye to highlight the positions of the maximum momentum components. \textbf{(b)} Fidelity between the exact two-excitation eigenstates, each of them indexed by a pair of quasi-momentum values ($k_1$,$k_2$), and the fermionized ansatz for $N$=50 qubits and $\kg  d/\pi=0.2$. } 
\label{fig2}
\end{centering}
\end{figure}

We numerically find the existence of two-excitation subradiant eigenstates $\ket{\psi_\xi^{(2)}}$, with a decay rate scaling as $\Gamma^{(2)}_{\xi}/\Gamma_{\rm 1D}\sim N^{-3}$ -- as in the single-excitation sector -- for the most subradiant eigenstates. These eigenstates reveal interesting properties in real and momentum space. One example is illustrated in the top of Fig.\,\ref{fig2}\,(a), where we consider the most subradiant two-excitation wavefunction $ \ket{\psi^{(2)}_{\xi=1}}=\sum_{m<n} c_{mn}|e_m,e_n\rangle$ for $k_{\rm 1D}d/\pi=0.2 $ 
and $N=20$ qubits, and plot both the probability amplitude $|c_{mn}|^2$ in real space (left) and $|c_{k_1,k_2}|^2$ in reciprocal space (right). Here, $c_{k_1,k_2}$ refers to the two-dimensional discrete Fourier transform of $c_{m n}$. In real space, the maximum in $|c_{mn}|^2$ occurs for $m\approx 6,n\approx 15$, revealing a tendency for the excitations to both repel each other, and avoid the system boundaries where they can be radiated. At the same time, in momentum space, a peak occurs around $k_{1,2}d/\pi=\pm 1$, coinciding with the dominant wavevectors $kd/\pi\approx \pm 1$ of the most subradiant single-excitation states [Fig.\,\ref{Fig_1}\,(c)]. 

A natural two-excitation wavefunction ansatz that realizes both the real- and momentum-space properties consists of taking an anti-symmetric combination of single-excitation eigenstates, which enforces a Pauli-like exclusion (``fermionization''). In particular, starting from the wavefunctions of the two most subradiant single-excitation eigenstates, we find that we can construct an accurate approximation of the most subradiant two-excitation eigenstate, 
\begin{equation} |\psi_{\xi=1}^{(F)}\rangle=\mathcal{N}\sum_{m<n} \left(c^{\xi=1}_m c^{\xi=2}_n-c^{\xi=2}_m c^{\xi=1}_n\right)|e_m, e_n\rangle\,, \end{equation} with $\mathcal{N}$ a normalization factor. For $k_{\rm 1D}d~{\rm mod}~\pi\neq 0$ and $k_{\rm 1D}d/\pi$ away from $0.5$, the $\xi=1,2$ single-excitation states have dominant wavevectors $(k_1,k_2)$ near the global decay rate minimum, e.g., at $k=\pi/d$ for $k_{\rm 1D}\,d/\pi=0.2$. For $k_{\rm 1D}\,d/\pi=0.5$, the fermionic ansatz also works well to describe the most subradiant two-excitation eigenstate (bottom of Fig.\,\ref{fig2}\,(a)). In this case, it is built from the most subradiant single-excitation eigenstates $k_1=\pi/d$ and $k_2=0$ (degenerate in decay rate), and results in the checkerboard pattern seen in the plot. 

To more generally examine the accuracy of the ansatz, we take the two-dimensional Fourier transform of each two-excitation eigenstate, and unambiguously assign a label of quasi-momentum indices $(k_1,k_2)$ to each state $|\psi^{(2)}_{(k_1,k_2)}\rangle$ based upon where the Fourier transform is peaked. We then compute the overlap fidelity $\mathcal{F}=|\langle \psi_{(k_1,k_2)}^{(F)}|\psi_{(k_1,k_2)}^{(2)}\rangle  |^2$ between the exact state and the fermionic ansatz composed of the single-excitation eigenstates $(k_1,k_2)$. As illustrated in Fig.\,\ref{fig2}\,(b), the ansatz works well when the two single-excitation states composing the eigenstate are strongly subradiant. In this case we find that the infidelity $1-\mathcal{F}$ scales with the qubit number as $1/N^2$ (see \,\ref{append_fansatz}). 
In the thermodynamic limit $N\rightarrow\infty$, we find that the decay rate of such subradiant ``fermionized" eigenstates approaches the sum of the decay rates of the single-excitation states they are composed of (see \ref{append_dadd}). \tbl{In the case of a 1D chain of atoms in \tblue{3D} free space, the fermionic ansatz was found to describe well both the most subradiant states \textit{and} the most radiant ones\,\cite{asenjo17}}. 

The conclusions made about the subradiant decay rate scaling and their fermionic nature -- exemplified here for two-excitations -- are found to extend to higher excitation numbers provided that the density of excitations is dilute: $m_{\rm ex} \ll N$. We now propose a procedure to observe this fermionic nature experimentally.

\section{Eigenstate preparation and measure of fermionic correlations\label{sec_3}}

\subsection{\tbl{Subradiant state preparation}}
To begin probing the fermionic character of two subradiant excitations, it would be desirable to generate a two-excitation Fock state.
It can be shown\,\cite{chang12} (see \ref{append_transfer}), that adding a single ancilla qubit to the array, which can be individually addressed, enables a collective Fock state with well-defined wavevector $k$ to be generated, by alternately creating an excitation in the ancilla and coherently transferring it to the array. The ancilla can subsequently be shifted far away in resonance frequency from the other qubits, so that it decouples from the dynamics under the effective Hamiltonian of Eq.\,(\ref{eq:Effective_Hamiltonian}). \tbl{To simplify the discussion we will assume for now that the preparation process leads to a perfect Fock state, and we address the role of imperfections in the subsequent section.}

Fock states $\ket{\varphi_k^{(m_{\rm ex})}}\sim (S_k^\dagger)^{m_{\rm ex}} \ket{g}^{\otimes N}$,  for low numbers of excitations $m_{\rm ex}$ and a $k$-vector corresponding to the decay rate minimum,  are found to have a significant overlap with the most subradiant eigenstates. For instance, when $k_{\rm 1D}d=0.7\pi$, the $N=10$ two-excitation state $\ket{\varphi_{k=0}^{(2)}}$ is found to have an overlap $\mathcal{F}_{\xi=1}^{(2)}=|\ex{\psi_\xi^{(2)}|\varphi_{k}^{(2)}}|^2\simeq 0.58$ with the most subradiant eigenstate (with only a weak dependence on the qubit number $N$).
\tbl{Moreover, the two-excitation state $\ket{\varphi_{k=0}^{(2)}}$ has an overlap of $\sum_{\xi_{\rm sr}}  \mathcal{F}_{\xi_{\rm sr}}^{(2)}\gtrsim 90\%$ with the entire subset of subradiant two-excitation eigenstates, where the summation captures all eigenstates with decay  \tblue{$\Gamma_{\xi_{\rm sr}}^{(2)}<2 \Gamma_{\rm 1D}$.} } 

\begin{figure}[tbh]
\begin{centering}
 \includegraphics[scale=0.7]{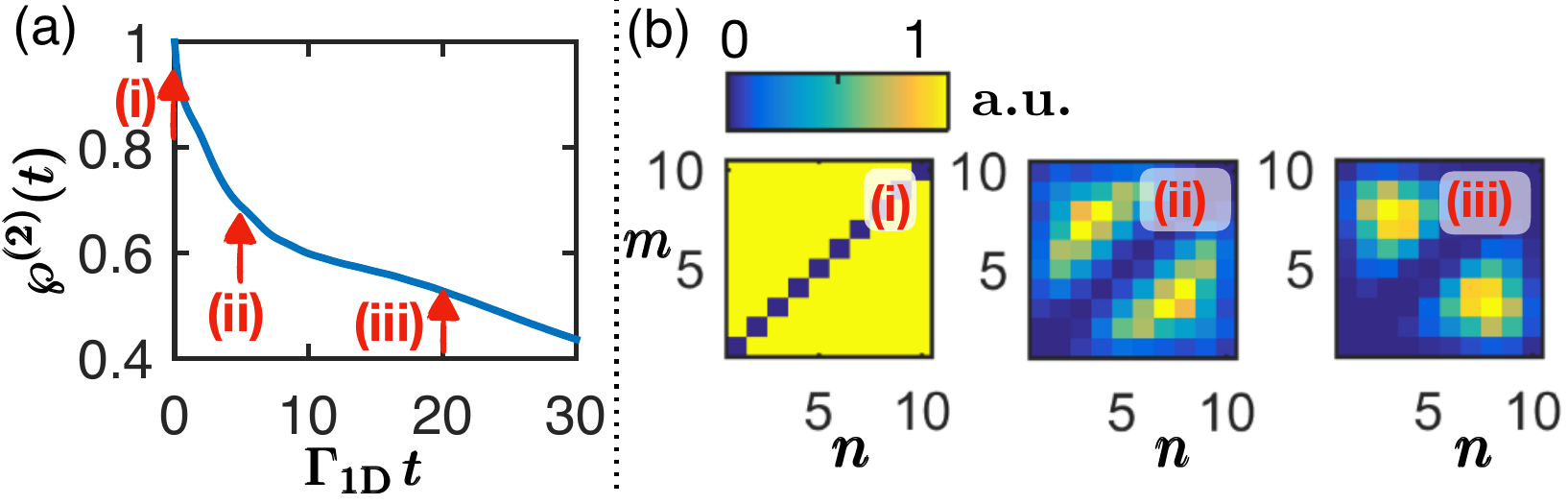}
\caption{Time evolution of the initial ($t=0$) state $\ket{\varphi_{k=0}^{(2)}}$ for $N=10$ qubits and $k_{\rm 1D}\,d=0.7\,\pi$. (a) Two-excitation probability $\wp^{(2)}$ in time. (b) Population of states $\ket{e_n,e_m}$ at the times $t$ as indicated by red arrows in (a). } 
\label{fig3}
\end{centering}
\end{figure}

\tbl{Here, we will show that interesting signatures of subradiance can be seen in time evolution, starting from $\ket{\varphi_{k=0}^{(2)}}$ as the initial state. We calculate the time evolution based on the master equation\,(\ref{master_eqn}), leading to a density matrix $\rho(t)$ in time. The probability for two excitations to remain in the system  $\wp^{(2)}(t)={\rm tr}(\mathcal{P}^{(2)}\rho(t) \mathcal{P}^{(2)})$, where $\mathcal{P}^{(2)}$ is the projector onto the atomic two-excitation subspace,  is depicted in Fig.\ref{fig3}\,(a). }

The majority of population persists for times $t\gg \Gamma_{\rm 1D}^{-1}$ due to subradiance. Furthermore, conditioned on finding two excitations in the system, the fermionic correlations increase in time as only the most subradiant states survive. \tbl{This is illustrated in Fig.\,\ref{fig3}\,(b) where the population of two-excitation states $\ex{e_n,e_m|\rho(t)|e_n,e_m}$ is shown for selected times. Fermionic} correlations are already evident at $\Gamma_{\rm 1D}t= 5$ (panel (ii)), where unconditionally $\sim 70\%$ of the original excitation remains\tbl{. At $\Gamma_{\rm 1D}t=20$ (panel (iii)) the state conditioned on two remaining excitations has nearly perfect overlap with the most subradiant state, and we find a fidelity $\mathcal{F}_{\xi =1}^{(2)}(t)=\ex{\psi_{\xi=1}^{(2)}|\rho(t)|\psi_{\xi=1}^{(2)}}/\wp^{(2)}(t)\gtrsim90\%$ (with $\wp^{(2)}(t) \sim0.5$).}

\tbl{\subsection{The role of imperfections}\label{sec_imperf}
In practice, both intrinsic decay through the waveguide, dephasing and excitation losses into channels other than the waveguide affect the probabilities and fidelities of  the excitation transfer and eigenstate convergence process. We model the non-waveguide decay (of rate $\Gamma'$) and dephasing (of rate $\gamma_d$) as uncorrelated and Markovian, that is 
\begin{equation} \mathcal{L}_{\rm dec}[\rho]=\frac{\Gamma'}{2}\sum_n \left[ 2\sigma_{ge}^n\rho\sigma_{eg}^n-\{\sigma_{ee}^n,\rho\}\right]+\gamma_d\,\sum_n \left[2\sigma_{ee}^n\rho\sigma_{ee}^n-\{\sigma_{ee}^n,\rho \}\right]  \end{equation}
where $\{\cdot,\cdot \}$ denotes the anticommutator.}

\begin{figure}[tbh!]
\begin{centering}
\includegraphics[scale=0.6]{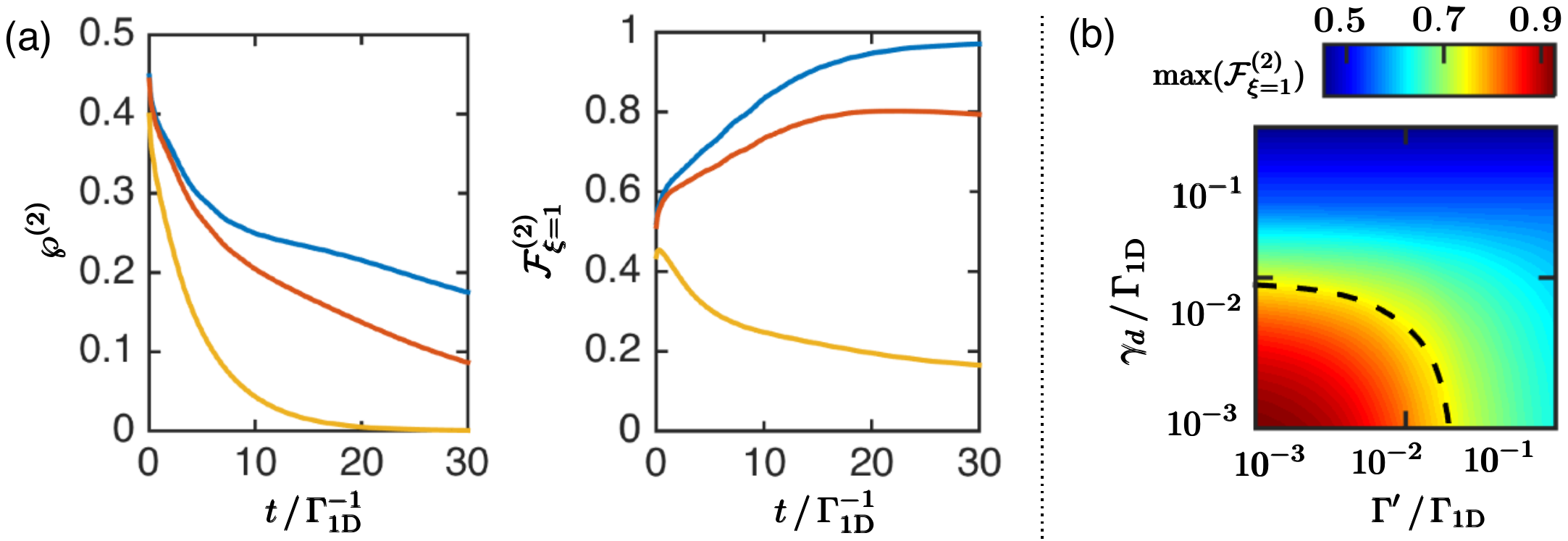}
\caption{\label{SM5fig} \tbl{Subradiant state preparation in the presence of imperfections for $N=10$ qubits and $k_{\rm 1D}\,d = 0.7\,\pi$.  (a)  Time evolution of the state obtained by an imperfect preparation of the Fock state $\ket{\varphi_{k=0}^{(2)}}$. Blue, red and orange lines correspond to dephasing rates (i) $\gamma_d=0$, (ii) $\gamma_d = 0.01\,\Gamma_{\rm 1D}$ and (iii) $\gamma_d=0.1\,\Gamma_{\rm 1D}$, respectively, with $\Gamma'=0$.  The state at time $t=0$ follows from the Fock state preparation sequence described in \ref{append_transfer}. The initial two-excitation probability $\wp^{(2)}(0)$ and fidelity with the target state $\mathcal{F}^{(2)}_{k=0}=\ex{\varphi_{k=0}^{(2)}|\rho(0)|\varphi_{k=0}^{(2)}}/\wp^{(2)}(0)$ are found to be (i) $\wp^{(2)}(0)=0.45$, $\mathcal{F}_{k=0}^{(2)}=0.99$, (ii) $\wp^{(2)}(0) = 0.44$, $\mathcal{F}^{(2)}_{k=0}=0.97$ and (iii) $\wp^{(2)}=0.4$, $\mathcal{F}^{(2)}_{k=0}=0.82$. The two-excitation probability $\wp^{(2)}(t)$ (left) and the fidelity $\mathcal{F}_{\xi =1}^{(2)}(t)=\ex{\psi_{\xi=1}^{(2)}|\rho(t)|\psi_{\xi=1}^{(2)}}/\wp^{(2)}(t)$ (right) in time are shown in the figure.} 
  (b)  Maximum fidelity ${\rm max}_t(\mathcal{F}_{\xi=1}^{(2)}(t))$ for preparing the most subradiant two-excitation eigenstate in the presence of additional loss and dephasing at rates $\Gamma'$ and $\gamma_d$, respectively. 
The maximization over the evolution time $t$ is conditioned on a probability of having two excitations in the system $\wp^{(2)}(t)\geq 0.2$. The dashed line marks a fidelity of 75\%.}
\end{centering}
\end{figure}

\tbl{The impact of imperfections on the Fock state $\ket{\varphi_k^{(2)}}$ preparation sequence is discussed in detail in \ref{append_transfer}. It turns out that for reasonably small imperfection rates $\Gamma',\,\gamma_d \ll \Gamma_{\rm 1D}$, the initial probability of creating two excitations is predominantly limited by the intrinsic decay $\Gamma_{\rm 1D}$, as we will shortly see. In particular in Fig.\,\ref{SM5fig}\,(a), we plot the time evolution of the two-excitation probability $\wp^{(2)}(t)$ and the overlap fidelity $\mathcal{F}_{\xi =1}^{(2)}(t)=\ex{\psi_{\xi=1}^{(2)}|{\rho}(t)|\psi_{\xi=1}^{(2)}}/\wp^{(2)}(t)$ between the most subradiant eigenstate and the system state conditioned on two excitations. The time $t=0$ corresponds to the state right after the Fock state preparation sequence, denoted as the state $\tilde{\rho}_{k=0}^{(2)}$ in \ref{append_B2}. The quantities are plotted for three different dephasing rates $\gamma_d=0$, $\gamma_d=0.01\,\Gamma_{\rm 1D}$ and $\gamma_d=0.1\,\Gamma_{\rm 1D}$. It can be seen that the initial two-excitation probabilities and fidelities $\wp^{(2)}(0)$ and $F_{\xi = 1}^{(2)}(0)$ are minimally affected by these dephasing values. In contrast, the probability is limited by $\Gamma_{\rm 1D}$ to $\wp^{(2)}(0)\simeq 0.45$, however, the state conditioned on two excitations can be shown to have a high overlap with the target state $\ket{\varphi_{k=0}^{(2)}}$ (with explicit values given in the caption to Fig.\,\ref{SM5fig}\,(a)). }

\tbl{An increase of fidelity $\mathcal{F}_\xi^{(2)}(t)$ in time is observed for dephasing rates $\gamma_d\leq 0.01\,\Gamma_{\rm 1D}$, whereas $\gamma_d = 0.1\,\Gamma_{\rm 1D}$ shows a decay of fidelity in time.  More generally, in order to obtain an increase of fidelity in time,  the dephasing rate $\gamma_d$ must be smaller than (or at least comparable to) the rate with which the most subradiant eigenstate is approached.}

\tbl{Moreover, we find that the probability decay $\wp^{(2)}(t)$, for realistic parameters of $\gamma_{d}, \Gamma'\sim 10^{-1}-10^{-3}\,\Gamma_{\rm 1D}$, to a good approximation only depends on the sum $\Gamma'+\gamma_{d}$. This suggests that dephasing essentially destroys subradiance and thus results in fast losses. On the other hand, the fidelity $\mathcal{F}^{(2)}_{\xi = 1}(t)$ only gets degraded by dephasing $\gamma_d$ and is independent of $\Gamma'$.}

\tbl{Combining both the preparation and time evolution, the maximum fidelity $\mathcal{F}^{(2)}_\xi$ (optimized over evolution time $t$) that can be achieved in the presence of loss and dephasing is illustrated in Fig.\,\ref{SM5fig}\,(b). To provide a more realistic experimental setting, in this plot we simultaneously require that the system has a non-negligible probability of at least $\wp^{(2)}(t)\geq 0.2$ to have two excitations in the system. A clear anti-bunching structure can be observed down to fidelities of around $75\%$ (marked by the dashed line in Fig.\,\ref{SM5fig}\,(b)), which limits $\Gamma', \gamma_d\lesssim10^{-2}\,\Gamma_{\rm 1D}$ for the $N=10$ qubit chain.}

\subsection{\tbl{Probing spatial correlations}}
Fermionic spatial correlations can be probed by using in-parallel readout of two resonators which are each dispersively coupled to their own qubit\,\cite{barends14,jeffrey14}. While finite readout time adds experimental difficulty to taking precise snapshots of spatial correlations in time, practical readout times of $100$~ns should be sufficient to capture dynamics on timescales of $5\Gamma_{\rm 1D}^{-1}\approx 800$~ns while maintaining $\Gamma'/\Gamma_{\rm 1D},\gamma_d/\Gamma_{\rm 1D}\approx 10^{-2}$. \tblue{This assumes uncorrelated relaxation and dephasing rates $\Gamma',\gamma_d\approx 2\pi\times 10$~kHz or coherence times of $\tau = 1/\gamma \approx 16~\mu$s. State-of-the-art superconducting qubit experiments in multiple groups have demonstrated coherence times on the order of or even in excess of this requirement\,\cite{barends14,reagor18,mckay17}.}

\section{Correlations in the emitted field} 

{Having discussed a possible scheme to observe interesting spatial correlations associated with multi-excitation subradiant states, we next discuss the photon correlations observable in their radiated fields.}
We first analyze what happens to the most subradiant eigenstate in the two-excitation sector,  $\ket{\psi_{\xi=1}^{(2)}}$, once a photon is emitted and detected, for example, on the left side of the chain. We find that the new conditional state after detecting a photon, $\ket{\psi_c}\sim E_L^+(t)\ket{\psi_{\xi=1}^{(2)}}$ is predominantly formed by a superposition of the two single-excitation states \tbl{that} $\ket{\psi_{\xi=1}^{(2)}}$ is composed of, i.e.,  $\ket{\psi_c}\simeq \alpha_1\ket{\psi^{(1)}_{\xi=1}}+ \alpha_2\ket{\psi^{(1)}_{\xi=2}}$.  More precisely, the projection of the conditional wavefunction onto any state besides the two most subradiant, $\varepsilon=1-|\alpha_1|^2-|\alpha_2|^2$, scales as $\varepsilon\sim1/N^2$ for most lattice constants $\kg d \neq 0.5\pi$.

After one photon is emitted at time $t$, the relative intensity of emission after a delay time $\tau$, normalized by the intensity at time $t$, is given by the two-photon correlation function 
 \begin{equation}
T^{(2)}(t,\tau)=\frac{\braket{\hat{E}_L^-(t)\hat{E}_L^-(t+\tau)\hat{E}_L^+(t+\tau)\hat{E}_L^+(t)}}{\braket{\hat{E}_L^-(t)\hat{E}_L^+(t)}^2}.\label{eq:T2}
\end{equation}
Prior experimental\,\cite{Menzel10,wallraff2011,Hoi12,houck2017} and theoretical\,\cite{blais2010,DiCandia14,Ramos17} work has demonstrated that such correlation functions can be measured in the microwave domain by amplifying the out-going photon field and performing correlation measurements between two linear detectors.

\begin{figure}[tbh]
\begin{centering}
\includegraphics[scale=0.7]{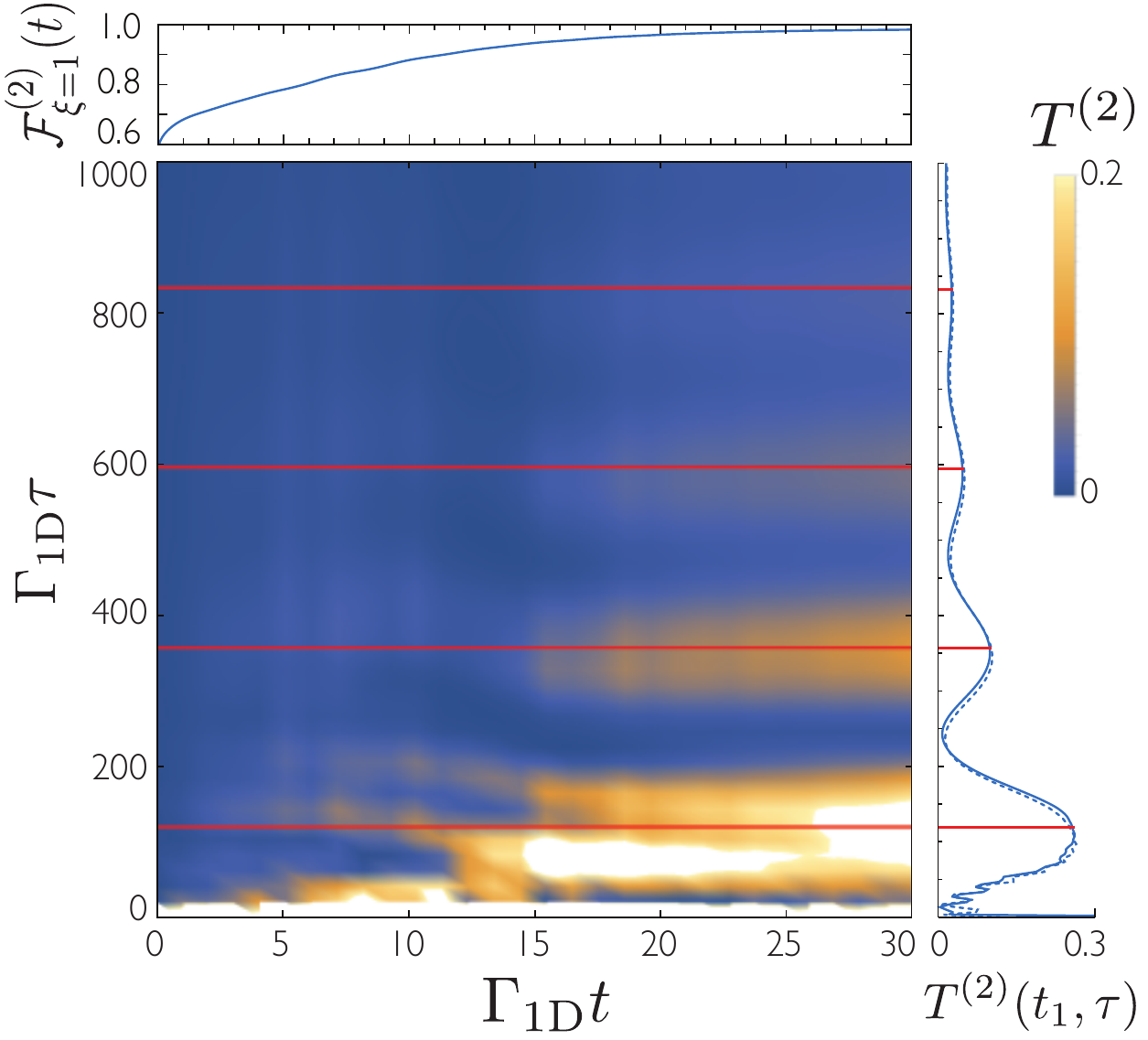}
\caption{Two-photon correlation function $T^{(2)}(t,\tau)$ for 10 qubits and $\kg d=0.7\pi$. At $t=0$, the qubits are prepared in the state $\ket{\varphi_{k=0}^{(2)}}=\mathcal{N}_2 (S_{k=0}^{\dagger})^2\ket{g}^{\otimes N}$. The red lines represent the delay times for which $T^{(2)}(t,\tau)$ is expected to be largest, i.e., $\tau_{\rm max}=n\pi/|J_{\xi=1}-J_{\xi=2}|$, with $n=\{1,3,5,7\}$. The plot on the right compares $T^{(2)}(t_1,\tau)$ at $t_1=30\Gamma_{\rm 1D}^{-1}$ for an initial state $\ket{\varphi_{k=0}^{(2)}}$ (solid curve) and $\ket{\psi_{\xi=1}^{(2)}}$ (dashed curve). The upper plot shows the evolution of the fidelity $\mathcal{F}_{\xi=1}^{(2)}$ with $t$.}\label{Fig4}
\end{centering}
\end{figure}

Figure~\ref{Fig4} shows $T^{(2)}(t,\tau)$ for a chain of 10 qubits with lattice constant $\kg d = 0.7\pi$. At $t=0$, the qubits are prepared in the state $\ket{\varphi_{k=0}^{(2)}}$. For short evolution times $t$, radiant state components lead to a rapid decrease of $T^{(2)}(t,\tau)$ with delay time $\tau$. At longer times $t$, when radiant components have largely vanished (see increasing overlap of the two-excitation subspace with the most subradiant state in the upper plot of Fig.\,\ref{Fig4}), a significant relative intensity can still remain at long delay times $\tau$. This leads to a visible emergence of oscillations in $T^{(2)}(t,\tau)$ as a function of $\tau$, coming from the interference in emission of the two single-excitation subradiant components (see the right part of Fig.~\ref{Fig4}). 
The oscillation period is determined by the difference in frequencies of the two most subradiant single excitation eigenstates, $J_{\xi=1}$ and $J_{\xi=2}$. In particular, the maxima in $T^{(2)}(t,\tau)$ occur at delay times $\tau_{\rm max}=n\pi/|J_{\xi=1}-J_{\xi=2}|$, with $n$ an odd integer. \tbl{In the presence of independent dephasing and decay, the oscillations in $T^{(2)}(t,\tau)$ can be observed provided that $\tau_{max}$ (for $n=1$) is shorter than the timescales of decay and dephasing. For the parameters of Fig. 6, this requires that $\gamma_d$, $\Gamma' \sim 10^{-3}\Gamma_{\rm 1D}$.}

\section{Conclusions}
In summary, we provided a comprehensive study of the subradiant properties of artificial atoms in waveguide QED, \tbl{which were found to bear close similarity to those of a 1D chain of atoms in \tblue{3D} free space despite the fact that the underlying Hamiltonians in these two systems differ considerably\,\cite{asenjo17}}. We have shown that this system represents an open quantum critical system with a closing of the Liouvillian gap in the thermodynamic limit. We have also shown that multi-excitation subradiant states exhibit ``fermionic'' spatial correlations, which can be probed in realistic experiments. 
This combination of features suggests that waveguide QED systems should be an attractive platform to broadly explore many-body open quantum systems, whose properties have drawn significant interest in recent years\,\cite{Bohnet16,luschen17,sieberer16,fossfeig17,henriet18}.

\ack
We are grateful to H.~J. Kimble for stimulating discussions. A.A.-G. was supported by an IQIM postdoctoral fellowship and the Global Marie Curie Fellowship LANTERN. P.B.D. was supported by a graduate fellowship from the Fannie and John Hertz Foundation. D.E.C. acknowledges support from Fundacio Privada Cellex, Spanish MINECO Severo Ochoa Program SEV-2015-0522, MINECO Plan Nacional Grant CANS, CERCA Programme/Generalitat de Catalunya, AGAUR Grant 2017 SGR 1334, and ERC Starting Grant FOQAL.

\appendix

\section{Multi-excitation subradiant states}
\subsection{Fidelity scaling of the fermionic ansatz}\label{append_fansatz}
As illustrated in Fig.\,\ref{fig2}\,(b) in the main text, highly subradiant two-excitation eigenstates $\ket{\psi_\xi^{(2)}}$ can often be well-approximated by an ansatz $\ket{\psi_\xi^{(F)}}$ that constructs fermionic combinations of single-excitation states.  Here, we analyze the scaling of the infidelity of such a fermionic ansatz $1-\mathcal{F}^{(2)}_\xi$, where $\mathcal{F}^{(2)}_\xi=|\ex{\psi_\xi^{(F)}|\psi_\xi^{(2)}}|^2$, with the qubit number $N$. The two-excitation eigenstates can be classified based on the associated wavevectors of their underlying single-excitation states. In particular, for the subradiant states considered here, these wavevectors take on values around $k=\pi/d$ and $k=0$, where the decay rates are lowest as shown in Fig.\,1\,(c) of the main text.  Two types of fermionic combinations can be distinguished: 
\begin{enumerate}[label=(\roman*)]
\item Those for which both single-excitation states can be associated with wavevectors corresponding to the same minimum (i.e. with both wavevectors around the decay rate minimum of either $k=0$ or $k=\pi/d$). 
\item Those that are combinations of both minima (i.e. one underlying single-excitation state associated with  $k=0$ and one with $k=\pi/d$).
\end{enumerate}
The first behavior can be seen in the most subradiant two-excitation eigenstate of $k_{\rm 1D}\,d=0.2\pi$ and the second for $k_{\rm 1D}\,d=0.5\pi$; both of these examples are pictured in Fig.\,\ref{fig2}\,(a) of the main text.

\begin{figure}[h!]
\begin{centering}
\includegraphics[scale=0.6]{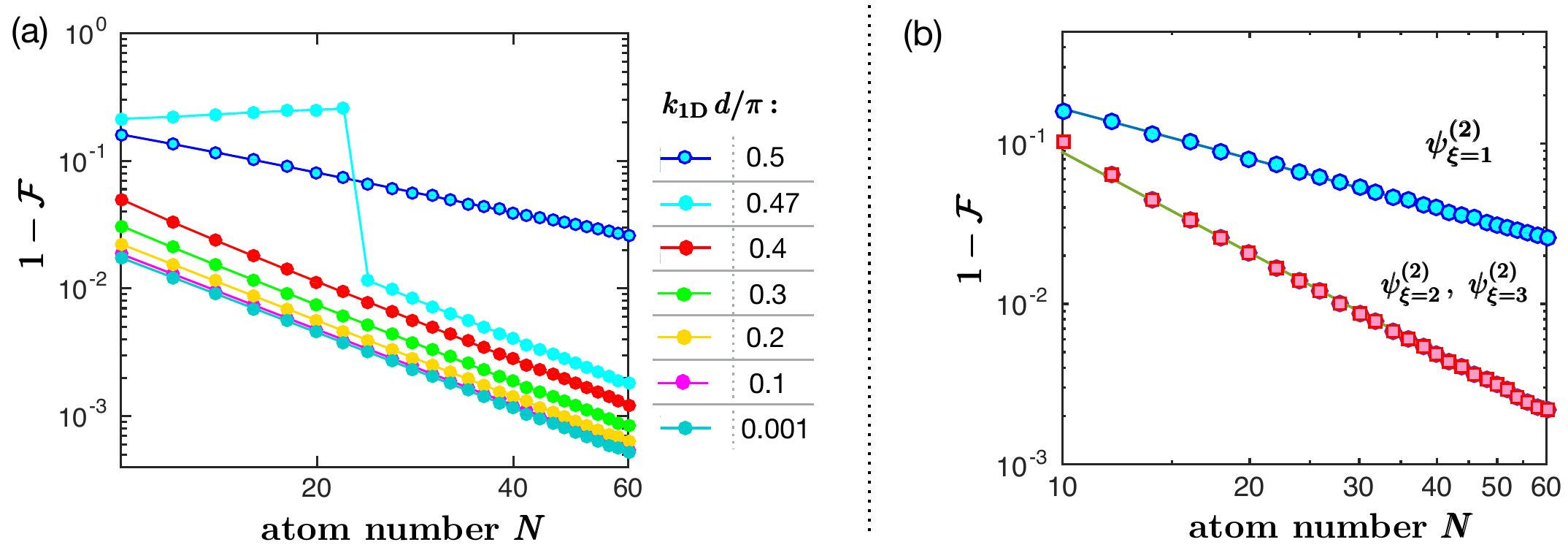}
\caption{\label{fdegscal}Infidelity $1-\mathcal{F}^{(2)}_\xi$ scaling for constructing two-excitation eigenstates as fermionic combinations of single excitation eigenstates. (a) Infidelity scaling  with the qubit number $N$ for the most subradiant eigenstate $\ket{\psi_{\xi=1}^{(2)}}$ and for selected values of $\kg \,d$. In the limit of large atom numbers a scaling $1-\mathcal{F}^{(2)}_\xi\sim N^{-s}$ can be identified with $s\simeq 1$ for $\kg \,d=0.5\pi$ and $s\simeq 2$ for all remaining $\kg \,d$. (b) Infidelity scaling with qubit number $N$ (for $\kg \,d=0.5\,\pi$) of the most subradiant two-excitation eigenstate $\ket{\psi_{\xi=1}^{(2)}}$ (blue, scaling: $s=1$) and the second- and third-most subradiant two-excitation eigenstates $\ket{\psi_{\xi=2}^{(2)}}$, $\ket{\psi_{\xi=3}^{(2)}}$ (red, scaling: $s=2$).}
\end{centering}
\end{figure}

The infidelity scaling with the atom number $N$, for describing the most subradiant two-excitation eigenstate by a fermionic ansatz, is depicted in Fig.\,\ref{fdegscal}\,(a). For $0<k_{\rm 1D}\,d<0.5\pi$, and for sufficiently large atom numbers, we find that fermionic combinations are of type (i), with an infidelity scaling $1-\mathcal{F}^{(2)}_\xi \sim N^{-2}$. Approaching $k_{\rm 1D}\,d = 0.5\pi$ (e.g., $k_{\rm 1D}\,d=0.47\pi$ in Fig.\,\ref{fdegscal}\,(a)), increasingly larger atom numbers are needed before the scaling property holds true. In that case, the combination of lowest-decaying single-excitation states would be of type (i), however, a combination of type (ii) forms the ansatz in the large $N$ limit (involving the lowest and third-lowest decaying eigenstates). The limiting cases $k_{\rm 1D}\,d = 0.5\pi$ and $k_{\rm 1D}\,d=0\,\pi$ are characterized by a singular behavior. For $k_{\rm 1D}\,d = 0.5\pi$, a configuration of type (ii) forms, with a distinct scaling $1-\mathcal{F}^{(2)}_\xi\sim N^{-1}$.
The case $k_{\rm 1D}\,d = 0$ represents the ``Dicke limit'', with many degenerate eigenstates of zero decay. Therefore, the unique definition of subradiant eigenstates and compositions breaks down.

In Fig.\,\ref{fdegscal}\,(b), the infidelity is plotted for the three most subradiant two-excitation eigenstates of $k_{\rm 1D}\,d=0.5\pi$. In these examples, only the most subradiant two-excitation eigenstate of $k_{\rm 1D}\,d=0.5\pi$ is of the type (ii) and thus shows a deviating infidelity scaling $\sim N^{-1}$ as opposed to the $\sim N^{-2}$ scaling of the remaining states.

\subsection{Scaling of decay rates}\label{append_dadd}

\begin{figure}[h!]
\begin{centering}
\includegraphics[scale=0.3]{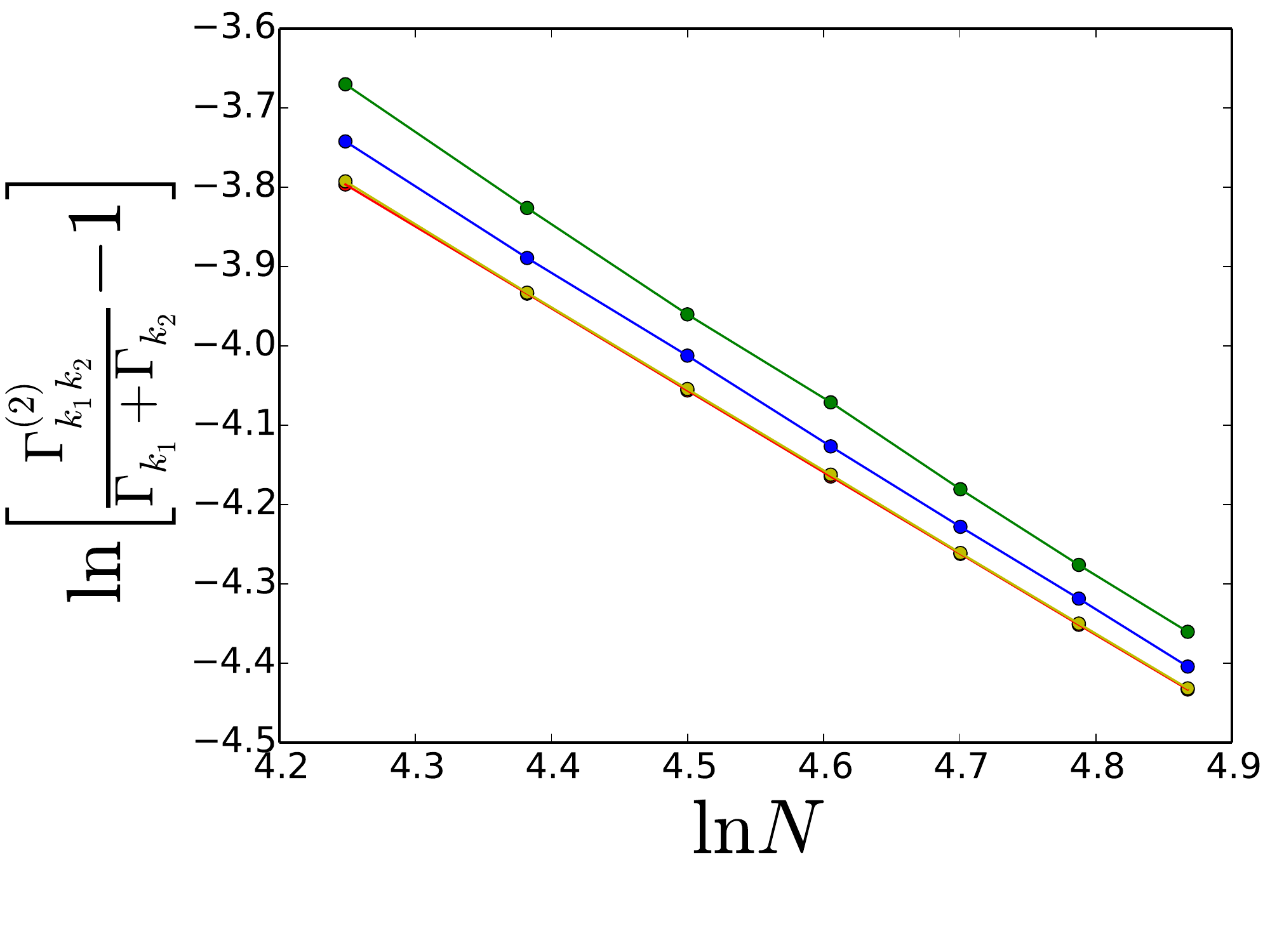}\includegraphics[scale=0.3]{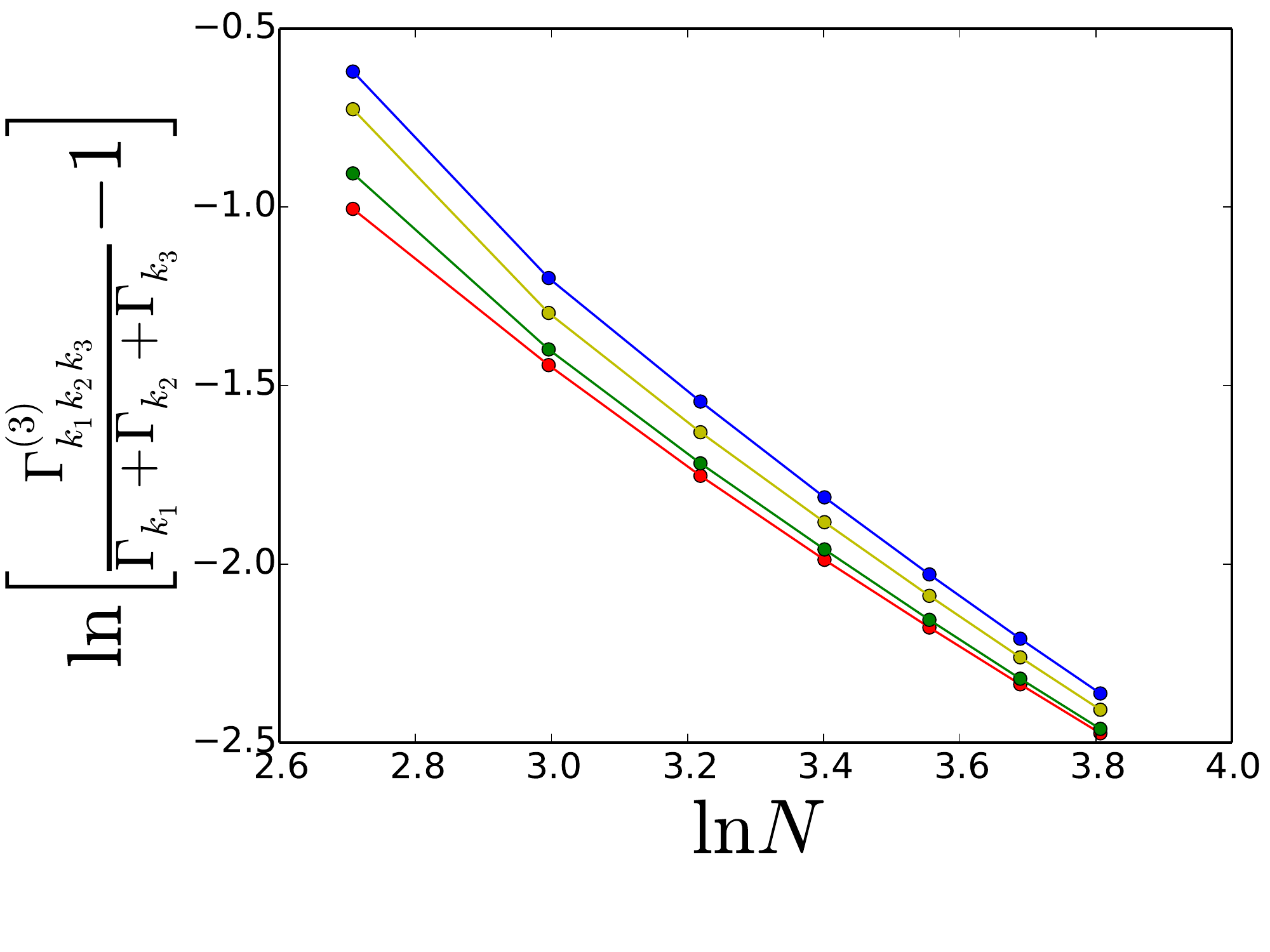}
\caption{\label{Fig_rate_scaling} Scaling of $r_2=\Gamma^{(2)}_{k_1,k_2}/(\Gamma_{k_1}+\Gamma_{k_2})-1$ and $r_3=\Gamma^{(3)}_{k_1,k_2,k_3}/(\Gamma_{k_1}+\Gamma_{k_2}+\Gamma_{k_3})-1$ in the limit of large atom number $N$. Each curve corresponds to one of the first four most subradiant eigenstates. We find that $r_2$ and $r_3$ evolve as $N^{-s}$ with $s\simeq 1$.  }
\end{centering}
\end{figure}

The most subradiant $m_{\rm ex}$-excitation eigenstates of the effective Hamiltonian are well described by the fermionic ansatz at low excitation densities ($m_{\rm ex} \ll N$) and can thus be written $|\psi^{(m_{\rm ex})}_{(k_1,k_2,..k_{m_{\rm ex}})}\rangle$, where $k_1,k_2,...,k_{m_{\rm ex}}$ denote the wavevectors of the single-excitation eigenstates composing this state. We find that such states have a decay rate $\Gamma^{(m_{\rm ex})}_{k_1,k_2,..,k_{m_{\rm ex}}}$ close to the sum of the decay rates of the single excitations they are composed of: $\Gamma^{(m_{\rm ex})}_{k_1,k_2,..,k_{m_{\rm ex}}}\sim\sum_{j=1}^{m_{\rm ex}} \Gamma_{k_j}$. In Fig.\,\ref{Fig_rate_scaling}, we illustrate this fact by showing that $r_{m_{\rm ex}}=\left(\Gamma^{(m_{\rm ex})}_{k_1,k_2,..,k_{m_{\rm ex}}}/\sum_{j=1}^{m_{\rm ex}} \Gamma_{k_j}\right)-1$ vanishes with $N$ for $m_{\rm ex}=2$ and $m_{\rm ex}=3$, for the first few most subradiant eigenstates.

\section{\tbl{Excitation transfer and Fock state preparation}}
In Sec.\,\ref{sec_3} of the main text, we used that a collective Fock state of definite wavevector can be prepared in the qubit chain configuration.  The preparation in such a state is achieved by adding an  ancilla qubit, which can be individually addressed and enables the transfer of excitations to the chain qubits. Subsequent to the preparation step, this ancilla can be effectively decoupled from the dynamics of the rest of the chain by shifting its frequency far out of resonance. Here we describe the preparation procedure and the distillation of subradiant eigenstates in more detail. \ref{append_transfer} introduces both the ancilla-chain configuration and the transfer protocol. The fidelities and success probabilities achieved by such a state preparation, in the presence of noise and imperfections, are discussed in \ref{append_B2}.

\subsection{Excitation transfer setup and protocol}\label{append_transfer}

\begin{figure}[tbh!]
\begin{centering}
\includegraphics[scale=0.6]{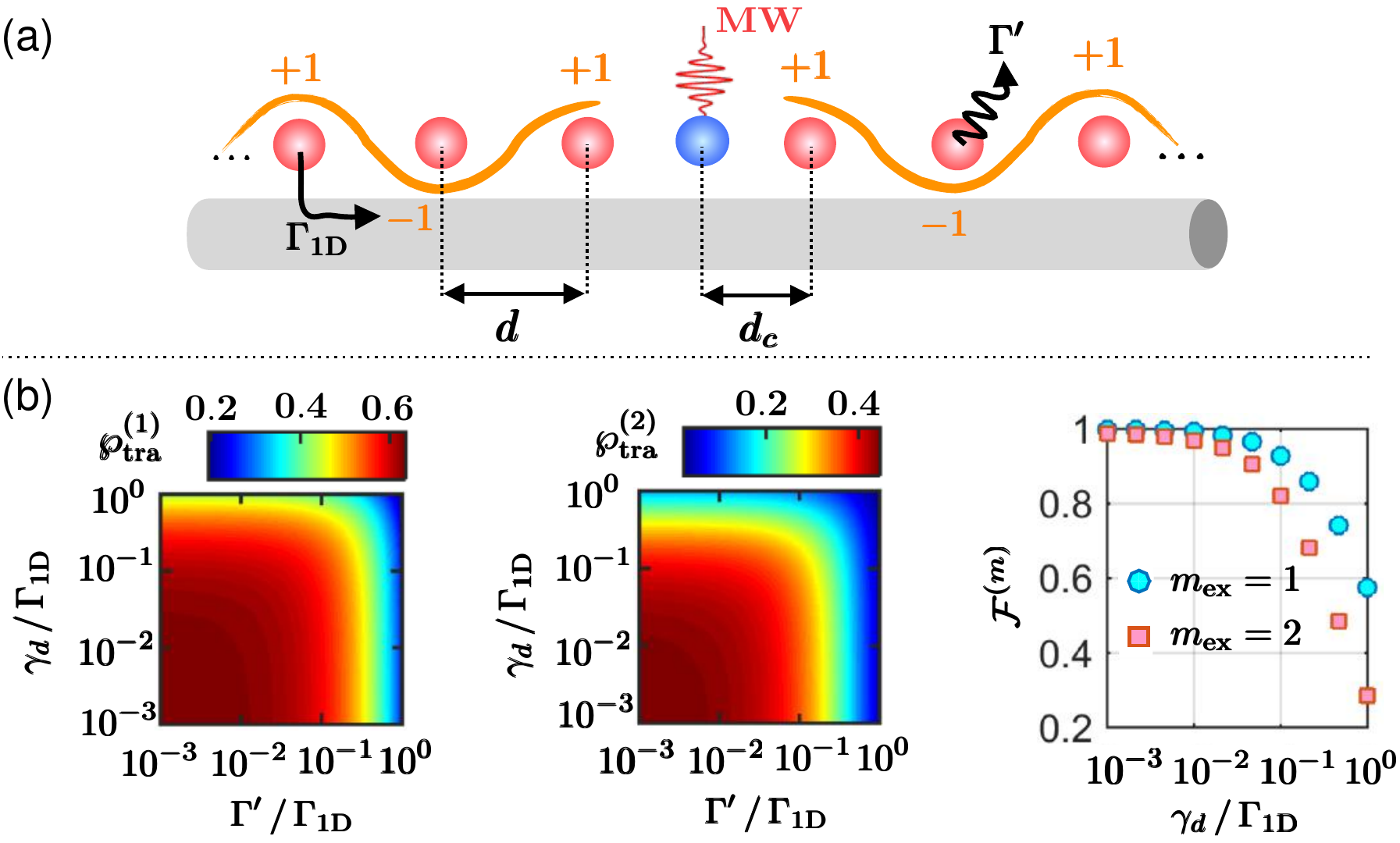}
\caption{\label{SM4fig} (a) Cavity configuration composed of two atomic ``mirror'' chains of $N/2$ qubits each (red circles) and a central ``cavity'' ancilla qubit (blue circle). A microwave excitation (MW) on the cavity qubit is coherently transfered to a collective excitation (orange, with numbers indicating the phase relation) of the mirror qubits. (b) Probability $\wp_{\rm tra}^{(m_{\rm ex})}$ (fidelity $\mathcal{F}^{(m_{\rm ex})}$  for the state $\varphi_{\rm mirr}^{(m_{\rm ex})}$) for the transfer of $m_{\rm ex}=1$ or $m_{\rm ex}=2$ excitations in the mirror configuration for $N=10$ chain qubits and various levels of free-space decay $\Gamma'$ and dephasing $\gamma_d$.  }
\end{centering}
\end{figure}

The transfer of excitations can be achieved in the so-called cavity configuration of waveguide QED\,\cite{chang12} illustrated in Fig.\,\ref{SM4fig}\,(a), where the $N$ chain qubits (depicted in red) form ``mirrors''   and an ancilla ``cavity'' qubit  (blue) is introduced at the midpoint. The chain qubits are equally spaced at a distance $d$ and the ancilla qubit is separated by $d_c$ from the nearest chain qubits. Specifically,  $k_{\rm 1D}\,d=\pi$ and  $k_{\rm 1D}\,d_c=\pi/2$, where $k_{\rm 1D}=\omega_{eg}/v$ is the wavevector of the qubit transition of frequency $\omega_{eg}$ within the waveguide of group velocity $v$.
The effective spin Hamiltonian for such a configuration, directly following from Eq.\,(\ref{eq:Effective_Hamiltonian}) in the main text, is given by\,\cite{chang12}
\begin{equation}\label{ham_mirr}  \mathcal{H}_{\rm \mathcal{C}} = \frac{\hbar\Gamma_{\rm 1D}}{2}\left[ (\sqrt{N}S_{\rm mirr}^\dagger\sigma_{\rm ge }^{\rm a}+{\rm h.c.})-i\left(NS^\dagger_{\rm rad}S_{\rm rad}+\sigma_{ee}^a\right)\right]  \end{equation}
with  $S_{\rm mirr/rad}^\dagger=\sum_{n=1}^{N/2}(-1)^n(\sigma_{\rm eg}^n\pm\sigma_{\rm eg}^{-n})/\sqrt{N}$ the collective transfer and decay operators (respectively) of the chain of mirror qubits. Here, $n$ and $-n$ enumerate the mirror qubits to the right and left of the ancilla qubit $a$, respectively. The first term in Eq.~(\ref{ham_mirr}) allows for a coherent exchange of excitations between the ancilla $a$ and the chain qubits, $\ket{e_a}\otimes (S_{\rm mirr}^\dagger)^{m_{\rm ex}-1}\ket{g}^{\otimes N}\Leftrightarrow \ket{g_a}\otimes  (S_{\rm mirr}^\dagger)^{m_{\rm ex}}\ket{g}^{\otimes N}$.
The resulting chain state exhibits zero (for $m_{\rm ex}=1$, $S_{\rm rad} S_{\rm mirr}^\dagger\ket{g}^{\otimes N}=0$) or low decay ($\sim \Gamma_{\rm 1D}/N$ for $1<m_{\rm ex}\ll N$). 

The preparation protocol, ideally resulting in a state $\ket{\varphi_{\rm mirr}^{(m_{\rm ex})}}=\mathcal{N} (S^\dagger_{\rm mirr})^{m_{\rm ex}}|g\rangle^{\otimes N}$, with $\mathcal{N}$ being a normalization constant, works as follows: Starting from all qubits in their ground state, a single-excitation Fock state in the mirrors can be prepared by applying a fast $\pi$-pulse to the ancilla qubit, $\ket{g_a}\rightarrow \ket{e_a}$, and subsequently waiting for a time $t_\pi^{(m_{\rm ex})}\simeq\pi/(\Gamma_{\rm 1D}\sqrt{N\,m_{\rm ex}})$ for that $m_{\rm ex}$-th (here $m_{\rm ex}=1$) excitation to be transfered to the mirrors. Higher number Fock states can be prepared by repeating the process.  Eliminating the ancilla qubit subsequent to the transfer, e.g., by detuning its frequency, reduces the system to a qubit chain periodically spaced by distance $d$. Note that the resulting state, if ideally prepared, automatically consists of a collective state where each qubit is excited with the same amplitude, and with a well-defined phase.

 Two further steps are needed to transform that Fock state to a state of definite wavevector on a qubit lattice of selected periodicity: First, a transformation  to the state  $\ket{\varphi_{k}^{(m_{\rm ex})}}\sim (S_k^\dagger)^{m_{\rm ex}}\,\ket{g}^{\otimes N}$ of wavevector $k$, which can be realized  by applying fast local phases to the qubits\,\cite{barends14} (here $S_k^\dagger=\sum_n e^{ikdn}\sigma_{eg}^n$). And  second, a dynamical modification of $k_{\rm 1D}\,d$ (in particular, via the resonant wavevector $k_{\rm 1D}$ itself), which can be accomplished by flux tuning the qubit transition frequency\,\cite{koch07}.

As discussed in the main text, a state $\ket{\varphi_{k}^{(m_{\rm ex})}}$ with $k$ a wavevector corresponding to the global decay minimum (e.g., $k=0$ for $k_{\rm 1D}\,d>0.5\pi$), is characterized by a significant overlap with the most subradiant $m_{\rm ex}$-excitation eigenstate. In particular, the state conditioned on $m_{\rm ex}$-excitations converges to the latter state in time, at the expense of a decreasing $m_{\rm ex}$-excitation probability.

\tbl{\subsection{Fock state preparation fidelity}\label{append_B2}
As outlined in Sect.\,\ref{sec_imperf} of the main text,  both intrinsic decay through the waveguide (at rate $\Gamma_{\rm 1D}$) and losses into other channels (at rate $\Gamma'$) as well as dephasing (at rate $\gamma_d$) limit the Fock state preparation protocol. Here, we analyze how these processes affect the transfer probability of the protocol disussed in the previous section. }

\tbl{Specifically, we consider the transfer of $m_{\rm ex}$ excitations, targeted towards ideally preparing the state $\ket{\varphi_{\rm mirr}^{(m_{\rm ex})}}$. }
For the simulations, the transfer times for each excitation $t_\pi^{(m_{\rm ex})}\approx \pi/(\Gamma_{\rm 1D}\sqrt{N\,m_{\rm ex}})$ have been optimized to maximize the fidelity with the state $\ket{\psi^{(m_{\rm ex})}}\sim\ket{g}_a\otimes (S_{\rm mirr}^\dagger)^{m_{\rm ex}}|g\rangle^{\otimes N} $, where the ancilla qubit $a$ is in the ground state and the chain qubits are in the target state. Moreover, we model the $\pi$-pulse to excite the ancilla qubit as an ideal gate that takes a negligible amount of time to perform.
 Incorporating loss and dephasing processes, the resulting chain state after the transfer, tracing out the ancilla qubit, is characterized by the reduced density matrix $\tilde{\rho}_{\rm mirr}^{(m_{\rm ex})}$. 

 The probability for ending up in an $m_{\rm ex}$-excitation state $\wp_{\rm tra}^{(m_{\rm ex})}$, equivalent to the trace of $\tilde{\rho}_{\rm mirr}^{(m_{\rm ex})}$ in the $m$-excitation subspace, and the fidelity of the $m_{\rm ex}$-excitation state $\mathcal{F}^{(m_{\rm ex})}=\ex{\varphi_{\rm mirr}^{(m_{\rm ex})}|\tilde{\rho}_{\rm mirr}^{(m_{\rm ex})}|\varphi_{\rm mirr}^{(m_{\rm ex})}}/\wp_{\rm tra}^{(m_{\rm ex})}$ are shown in Fig.\,\ref{SM4fig}\,(b) for $N=10$ qubits as a function of $\Gamma'$ and $\gamma_d$.  Note that even for $\Gamma'=\gamma_d=0$, the two-excitation transfer probability (fidelity) is limited to $\wp^{(2)}_{\rm tra}\simeq 0.45$ ($\mathcal{F}^{(2)}\simeq0.99$) by the loss processes through the waveguide. However, as the dominant loss mechanism stems from the emission of the ancilla qubit into the waveguide, decreasing the ancilla-waveguide coupling compared to the waveguide coupling of the chain qubits allows for an increased fidelity and transfer probability at the expense of a longer transfer time. The fidelity only depends on the dephasing rate $\gamma_d$ and is independent of $\Gamma'$.

The state $\tilde{\rho}_{\rm mirr}^{(m_{\rm ex})}$ resulting from the excitation transfer ideally excites the qubits with the well-defined phases illustrated in Fig.\,\ref{SM4fig}\,(a) -- the qubits in the left and right ``mirrors'' each have alternating phases corresponding to a spin wavevector of $kd=\pi$, while a ``phase slip" occurs between the left and right mirrors as the qubits closest to the ancilla have the same phase. We want to subsequently convert this state to a state of well-defined $k$, here assumed to be $k=0$, in the attempt to ideally prepare the $m$-excitation state $\ket{\varphi_{k=0}^{(m_{\rm ex})}}$. We do so by the phase adjustment operation: $\tilde{\rho}^{(m_{\rm ex})}_{k=0}=S_\pi\tilde{\rho}_{\rm mirr}^{(m_{\rm ex})}S_\pi^\dagger$, where $S_\pi=\prod_{n}\exp(-i[1-(-1)^n]\pi(\sigma_{ee}^n+\sigma_{ee}^{-n})/2)$, with $n$ and $-n$ enumerating the left and right mirror qubits, respectively. As for the $\pi$-pulse operation, we assumed this gate to be instantaneous and of unit fidelity.

\clearpage
\section*{References}
\bibliography{refs}

\begin{thebibliography}{10}
\providecommand{\url}[1]{\texttt{#1}}
\providecommand{\urlprefix}{URL }
\providecommand{\bibinfo}[2]{#2}
\providecommand{\eprint}[2][]{\url{#2}}

\bibitem{astafiev10}
\bibinfo{author}{Astafiev, O.}, \bibinfo{author}{Zagoskin, A.~M.},
  \bibinfo{author}{Abdumalikov, A.~A.}, \bibinfo{author}{Pashkin, Y.~A.},
  \bibinfo{author}{Yamamoto, T.}, \bibinfo{author}{Inomata, K.},
  \bibinfo{author}{Nakamura, Y.} \& \bibinfo{author}{Tsai, J.~S.}
\newblock
\bibinfo{title}{Resonance fluorescence of a single artificial atom}.
  \href{http://dx.doi.org/10.1126/science.1181918}{\newblock
  \emph{\bibinfo{journal}{Science}} \textbf{\bibinfo{volume}{327}},
  \bibinfo{pages}{840--843} (\bibinfo{year}{2010})}.

\bibitem{wQED_circuit_2}
\bibinfo{author}{Hoi, I.-C.}, \bibinfo{author}{Wilson, C.~M.},
  \bibinfo{author}{Johansson, G.}, \bibinfo{author}{Palomaki, T.},
  \bibinfo{author}{Peropadre, B.} \& \bibinfo{author}{Delsing, P.}
\newblock
\bibinfo{title}{Demonstration of a single-photon router in the microwave
  regime}. \href{http://dx.doi.org/10.1103/PhysRevLett.107.073601}{\newblock
  \emph{\bibinfo{journal}{Phys. Rev. Lett.}} \textbf{\bibinfo{volume}{107}},
  \bibinfo{pages}{073601} (\bibinfo{year}{2011})}.

\bibitem{van_Loo_Science}
\bibinfo{author}{van Loo, A.~F.}, \bibinfo{author}{Fedorov, A.},
  \bibinfo{author}{Lalumi{\`e}re, K.}, \bibinfo{author}{Sanders, B.~C.},
  \bibinfo{author}{Blais, A.} \& \bibinfo{author}{Wallraff, A.}
\newblock
\bibinfo{title}{Photon-mediated interactions between distant artificial atoms}.
  \href{http://dx.doi.org/10.1126/science.1244324}{\newblock
  \emph{\bibinfo{journal}{Science}} \textbf{\bibinfo{volume}{342}},
  \bibinfo{pages}{1494--1496} (\bibinfo{year}{2013})}.

\bibitem{chang07}
\bibinfo{author}{Chang, D.~E.}, \bibinfo{author}{Sorensen, A.~S.},
  \bibinfo{author}{Demler, E.~A.} \& \bibinfo{author}{Lukin, M.~D.}
\newblock
\bibinfo{title}{A single-photon transistor using nanoscale surface plasmons}.
  \href{http://dx.doi.org/10.1038/nphys708}{\newblock
  \emph{\bibinfo{journal}{Nature Phys.}} \textbf{\bibinfo{volume}{3}},
  \bibinfo{pages}{807--812} (\bibinfo{year}{2007})}.

\bibitem{Shen07}
\bibinfo{author}{Shen, J.-T.} \& \bibinfo{author}{Fan, S.}
\newblock
\bibinfo{title}{Strongly correlated two-photon transport in a one-dimensional
  waveguide coupled to a two-level system}.
  \href{http://dx.doi.org/10.1103/PhysRevLett.98.153003}{\newblock
  \emph{\bibinfo{journal}{Phys. Rev. Lett.}} \textbf{\bibinfo{volume}{98}},
  \bibinfo{pages}{153003} (\bibinfo{year}{2007})}.

\bibitem{vetsch04}
\bibinfo{author}{Vetsch, E.}, \bibinfo{author}{Reitz, D.},
  \bibinfo{author}{Sagu\'e, G.}, \bibinfo{author}{Schmidt, R.},
  \bibinfo{author}{Dawkins, S.~T.} \& \bibinfo{author}{Rauschenbeutel, A.}
\newblock
\bibinfo{title}{Optical interface created by laser-cooled atoms trapped in the
  evanescent field surrounding an optical nanofiber}.
  \href{http://dx.doi.org/10.1103/PhysRevLett.104.203603}{\newblock
  \emph{\bibinfo{journal}{Phys. Rev. Lett.}} \textbf{\bibinfo{volume}{104}},
  \bibinfo{pages}{203603} (\bibinfo{year}{2010})}.

\bibitem{zheng10}
\bibinfo{author}{Zheng, H.}, \bibinfo{author}{Gauthier, D.~J.} \&
  \bibinfo{author}{Baranger, H.~U.}
\newblock
\bibinfo{title}{Waveguide {QED}: {Many-body} bound-state effects in coherent
  and {Fock-state} scattering from a two-level system}.
  \href{http://dx.doi.org/10.1103/PhysRevA.82.063816}{\newblock
  \emph{\bibinfo{journal}{Phys. Rev. A}} \textbf{\bibinfo{volume}{82}},
  \bibinfo{pages}{063816} (\bibinfo{year}{2010})}.

\bibitem{pichlerPRA15}
\bibinfo{author}{Pichler, H.}, \bibinfo{author}{Ramos, T.},
  \bibinfo{author}{Daley, A.~J.} \& \bibinfo{author}{Zoller, P.}
\newblock
\bibinfo{title}{Quantum optics of chiral spin networks}.
  \href{http://dx.doi.org/10.1103/PhysRevA.91.042116}{\newblock
  \emph{\bibinfo{journal}{Phys. Rev. A}} \textbf{\bibinfo{volume}{91}},
  \bibinfo{pages}{042116} (\bibinfo{year}{2015})}.

\bibitem{sanchez-burillo17}
\bibinfo{author}{S\'anchez-Burillo, E.}, \bibinfo{author}{Zueco, D.},
  \bibinfo{author}{Mart\'{\i}n-Moreno, L.} \&
  \bibinfo{author}{Garc\'{\i}a-Ripoll, J.~J.}
\newblock
\bibinfo{title}{Dynamical signatures of bound states in waveguide {QED}}.
  \href{http://dx.doi.org/10.1103/PhysRevA.96.023831}{\newblock
  \emph{\bibinfo{journal}{Phys. Rev. A}} \textbf{\bibinfo{volume}{96}},
  \bibinfo{pages}{023831} (\bibinfo{year}{2017})}.

\bibitem{Roy17}
\bibinfo{author}{Roy, D.}, \bibinfo{author}{Wilson, C.~M.} \&
  \bibinfo{author}{Firstenberg, O.}
\newblock
\bibinfo{title}{Colloquium: {Strongly} interacting photons in one-dimensional
  continuum}. \href{http://dx.doi.org/10.1103/RevModPhys.89.021001}{\newblock
  \emph{\bibinfo{journal}{Rev. Mod. Phys.}} \textbf{\bibinfo{volume}{89}},
  \bibinfo{pages}{021001} (\bibinfo{year}{2017})}.

\bibitem{GU17}
\bibinfo{author}{Gu, X.}, \bibinfo{author}{Kockum, A.~F.},
  \bibinfo{author}{Miranowicz, A.}, \bibinfo{author}{xi~Liu, Y.} \&
  \bibinfo{author}{Nori, F.}
\newblock
\bibinfo{title}{Microwave photonics with superconducting quantum circuits}.
  \href{http://dx.doi.org/10.1016/j.physrep.2017.10.002}{\newblock
  \emph{\bibinfo{journal}{Phys. Rep.}} \textbf{\bibinfo{volume}{718-719}},
  \bibinfo{pages}{1 -- 102} (\bibinfo{year}{2017})}.

\bibitem{shen05}
\bibinfo{author}{Shen, J.-T.} \& \bibinfo{author}{Fan, S.}
\newblock
\bibinfo{title}{Coherent single photon transport in a one-dimensional waveguide
  coupled with superconducting quantum bits}.
  \href{http://dx.doi.org/10.1103/PhysRevLett.95.213001}{\newblock
  \emph{\bibinfo{journal}{Phys. Rev. Lett.}} \textbf{\bibinfo{volume}{95}},
  \bibinfo{pages}{213001} (\bibinfo{year}{2005})}.

\bibitem{abdumalikov10}
\bibinfo{author}{Abdumalikov, A.~A.}, \bibinfo{author}{Astafiev, O.},
  \bibinfo{author}{Zagoskin, A.~M.}, \bibinfo{author}{Pashkin, Y.~A.},
  \bibinfo{author}{Nakamura, Y.} \& \bibinfo{author}{Tsai, J.~S.}
\newblock
\bibinfo{title}{Electromagnetically induced transparency on a single artificial
  atom}. \href{http://dx.doi.org/10.1103/PhysRevLett.104.193601}{\newblock
  \emph{\bibinfo{journal}{Phys. Rev. Lett.}} \textbf{\bibinfo{volume}{104}},
  \bibinfo{pages}{193601} (\bibinfo{year}{2010})}.

\bibitem{Hoi13}
\bibinfo{author}{Hoi, I.-C.}, \bibinfo{author}{Wilson, C.~M.},
  \bibinfo{author}{Johansson, G.}, \bibinfo{author}{Lindkvist, J.},
  \bibinfo{author}{Peropadre, B.}, \bibinfo{author}{Palomaki, T.} \&
  \bibinfo{author}{Delsing, P.}
\newblock
\bibinfo{title}{Microwave quantum optics with an artificial atom in
  one-dimensional open space}.
  \href{http://dx.doi.org/10.1088/1367-2630/15/2/025011}{\newblock
  \emph{\bibinfo{journal}{New J. Phys.}} \textbf{\bibinfo{volume}{15}},
  \bibinfo{pages}{025011} (\bibinfo{year}{2013})}.

\bibitem{Wilson2011}
\bibinfo{author}{Wilson, C.~M.}, \bibinfo{author}{Johansson, G.},
  \bibinfo{author}{Pourkabirian, A.}, \bibinfo{author}{Simoen, M.},
  \bibinfo{author}{Johansson, J.~R.}, \bibinfo{author}{Duty, T.},
  \bibinfo{author}{Nori, F.} \& \bibinfo{author}{Delsing, P.}
\newblock
\bibinfo{title}{{Observation of the dynamical Casimir effect in a
  superconducting circuit}}.
  \href{http://dx.doi.org/10.1038/nature10561}{\newblock
  \emph{\bibinfo{journal}{Nature}} \textbf{\bibinfo{volume}{479}},
  \bibinfo{pages}{376--379} (\bibinfo{year}{2011})}.

\bibitem{Hoi15}
\bibinfo{author}{Hoi, I.~C.}, \bibinfo{author}{Kockum, A.~F.},
  \bibinfo{author}{Tornberg, L.}, \bibinfo{author}{Pourkabirian, A.},
  \bibinfo{author}{Johansson, G.}, \bibinfo{author}{Delsing, P.} \&
  \bibinfo{author}{Wilson, C.~M.}
\newblock
\bibinfo{title}{Probing the quantum vacuum with an artificial atom in front of
  a mirror}. \href{http://dx.doi.org/10.1038/nphys3484}{\newblock
  \emph{\bibinfo{journal}{Nature Phys.}} \textbf{\bibinfo{volume}{11}},
  \bibinfo{pages}{1045 EP --} (\bibinfo{year}{2015})}.

\bibitem{tudela13}
\bibinfo{author}{Gonz\'alez-Tudela, A.} \& \bibinfo{author}{Porras, D.}
\newblock
\bibinfo{title}{Mesoscopic entanglement induced by spontaneous emission in
  solid-state quantum optics}.
  \href{http://dx.doi.org/10.1103/PhysRevLett.110.080502}{\newblock
  \emph{\bibinfo{journal}{Phys. Rev. Lett.}} \textbf{\bibinfo{volume}{110}},
  \bibinfo{pages}{080502} (\bibinfo{year}{2013})}.

\bibitem{Fang15}
\bibinfo{author}{Fang, Y.-L.~L.} \& \bibinfo{author}{Baranger, H.~U.}
\newblock
\bibinfo{title}{Waveguide {QED}: {Power} spectra and correlations of two
  photons scattered off multiple distant qubits and a mirror}.
  \href{http://dx.doi.org/10.1103/PhysRevA.91.053845}{\newblock
  \emph{\bibinfo{journal}{Phys. Rev. A}} \textbf{\bibinfo{volume}{91}},
  \bibinfo{pages}{053845} (\bibinfo{year}{2015})}.

\bibitem{chang12}
\bibinfo{author}{Chang, D.~E.}, \bibinfo{author}{Jiang, L.},
  \bibinfo{author}{Gorshkov, A.~V.} \& \bibinfo{author}{Kimble, H.~J.}
\newblock
\bibinfo{title}{Cavity {QED} with atomic mirrors}.
  \href{http://dx.doi.org/10.1088/1367-2630/14/6/063003}{\newblock
  \emph{\bibinfo{journal}{New J. Phys.}} \textbf{\bibinfo{volume}{14}},
  \bibinfo{pages}{063003} (\bibinfo{year}{2012})}.

\bibitem{Tudela17}
\bibinfo{author}{Gonz\'alez-Tudela, A.}, \bibinfo{author}{Paulisch, V.},
  \bibinfo{author}{Kimble, H.~J.} \& \bibinfo{author}{Cirac, J.~I.}
\newblock
\bibinfo{title}{Efficient multiphoton generation in waveguide quantum
  electrodynamics}.
  \href{http://dx.doi.org/10.1103/PhysRevLett.118.213601}{\newblock
  \emph{\bibinfo{journal}{Phys. Rev. Lett.}} \textbf{\bibinfo{volume}{118}},
  \bibinfo{pages}{213601} (\bibinfo{year}{2017})}.

\bibitem{dzsotjan2010}
\bibinfo{author}{Dzsotjan, D.}, \bibinfo{author}{S\o{}rensen, A.~S.} \&
  \bibinfo{author}{Fleischhauer, M.}
\newblock
\bibinfo{title}{Quantum emitters coupled to surface plasmons of a nanowire: {A}
  {Green's} function approach}.
  \href{http://dx.doi.org/10.1103/PhysRevB.82.075427}{\newblock
  \emph{\bibinfo{journal}{Phys. Rev. B}} \textbf{\bibinfo{volume}{82}},
  \bibinfo{pages}{075427} (\bibinfo{year}{2010})}.

\bibitem{Paulisch16}
\bibinfo{author}{Paulisch, V.}, \bibinfo{author}{Kimble, H.~J.} \&
  \bibinfo{author}{Gonz{\'a}lez-Tudela, A.}
\newblock
\bibinfo{title}{Universal quantum computation in waveguide {QED} using
  decoherence free subspaces}.
  \href{http://dx.doi.org/10.1088/1367-2630/18/4/043041}{\newblock
  \emph{\bibinfo{journal}{New J. Phys.}} \textbf{\bibinfo{volume}{18}},
  \bibinfo{pages}{043041} (\bibinfo{year}{2016})}.

\bibitem{Haakh2016}
\bibinfo{author}{Haakh, H.~R.}, \bibinfo{author}{Faez, S.} \&
  \bibinfo{author}{Sandoghdar, V.}
\newblock
\bibinfo{title}{Polaritonic normal-mode splitting and light localization in a
  one-dimensional nanoguide}.
  \href{http://dx.doi.org/10.1103/PhysRevA.94.053840}{\newblock
  \emph{\bibinfo{journal}{Phys. Rev. A}} \textbf{\bibinfo{volume}{94}},
  \bibinfo{pages}{053840} (\bibinfo{year}{2016})}.

\bibitem{asenjo17}
\bibinfo{author}{Asenjo-Garcia, A.}, \bibinfo{author}{Moreno-Cardoner, M.},
  \bibinfo{author}{Albrecht, A.}, \bibinfo{author}{Kimble, H.~J.} \&
  \bibinfo{author}{Chang, D.~E.}
\newblock
\bibinfo{title}{Exponential improvement in photon storage fidelities using
  subradiance and ``selective radiance'' in atomic arrays}.
  \href{http://dx.doi.org/10.1103/PhysRevX.7.031024}{\newblock
  \emph{\bibinfo{journal}{Phys. Rev. X}} \textbf{\bibinfo{volume}{7}},
  \bibinfo{pages}{031024} (\bibinfo{year}{2017})}.

\bibitem{koch07}
\bibinfo{author}{Koch, J.} \emph{et~al.}
\newblock
\bibinfo{title}{Charge-insensitive qubit design derived from the {C}ooper pair
  box}. \href{http://dx.doi.org/10.1103/PhysRevA.76.042319}{\newblock
  \emph{\bibinfo{journal}{Phys. Rev. A}} \textbf{\bibinfo{volume}{76}},
  \bibinfo{pages}{042319} (\bibinfo{year}{2007})}.

\bibitem{Lalumiere_PRA}
\bibinfo{author}{Lalumi\`ere, K.}, \bibinfo{author}{Sanders, B.~C.},
  \bibinfo{author}{van Loo, A.~F.}, \bibinfo{author}{Fedorov, A.},
  \bibinfo{author}{Wallraff, A.} \& \bibinfo{author}{Blais, A.}
\newblock
\bibinfo{title}{Input-output theory for waveguide {QED} with an ensemble of
  inhomogeneous atoms}.
  \href{http://dx.doi.org/10.1103/PhysRevA.88.043806}{\newblock
  \emph{\bibinfo{journal}{Phys. Rev. A}} \textbf{\bibinfo{volume}{88}},
  \bibinfo{pages}{043806} (\bibinfo{year}{2013})}.

\bibitem{Tommaso_NJP_2015}
\bibinfo{author}{Caneva, T.}, \bibinfo{author}{Manzoni, M.~T.},
  \bibinfo{author}{Shi, T.}, \bibinfo{author}{Douglas, J.~S.},
  \bibinfo{author}{Cirac, J.~I.} \& \bibinfo{author}{Chang, D.~E.}
\newblock
\bibinfo{title}{Quantum dynamics of propagating photons with strong
  interactions: a generalized input--output formalism}.
  \href{http://dx.doi.org/10.1088/1367-2630/17/11/113001}{\newblock
  \emph{\bibinfo{journal}{New J. Phys.}} \textbf{\bibinfo{volume}{17}},
  \bibinfo{pages}{113001} (\bibinfo{year}{2015})}.

\bibitem{dicke54}
\bibinfo{author}{Dicke, R.~H.}
\newblock
\bibinfo{title}{Coherence in spontaneous radiation processes}.
  \href{http://dx.doi.org/10.1103/PhysRev.93.99}{\newblock
  \emph{\bibinfo{journal}{Phys. Rev.}} \textbf{\bibinfo{volume}{93}},
  \bibinfo{pages}{99--110} (\bibinfo{year}{1954})}.

\bibitem{corzo16}
\bibinfo{author}{Corzo, N.~V.}, \bibinfo{author}{Gouraud, B.},
  \bibinfo{author}{Chandra, A.}, \bibinfo{author}{Goban, A.},
  \bibinfo{author}{Sheremet, A.~S.}, \bibinfo{author}{Kupriyanov, D.~V.} \&
  \bibinfo{author}{Laurat, J.}
\newblock
\bibinfo{title}{Large {Bragg} reflection from one-dimensional chains of trapped
  atoms near a nanoscale waveguide}.
  \href{http://dx.doi.org/10.1103/PhysRevLett.117.133603}{\newblock
  \emph{\bibinfo{journal}{Phys. Rev. Lett.}} \textbf{\bibinfo{volume}{117}},
  \bibinfo{pages}{133603} (\bibinfo{year}{2016})}.

\bibitem{sorensen16}
\bibinfo{author}{S\o{}rensen, H.~L.}, \bibinfo{author}{B\'eguin, J.-B.},
  \bibinfo{author}{Kluge, K.~W.}, \bibinfo{author}{Iakoupov, I.},
  \bibinfo{author}{S\o{}rensen, A.~S.}, \bibinfo{author}{M\"uller, J.~H.},
  \bibinfo{author}{Polzik, E.~S.} \& \bibinfo{author}{Appel, J.}
\newblock
\bibinfo{title}{Coherent backscattering of light off one-dimensional atomic
  strings}. \href{http://dx.doi.org/10.1103/PhysRevLett.117.133604}{\newblock
  \emph{\bibinfo{journal}{Phys. Rev. Lett.}} \textbf{\bibinfo{volume}{117}},
  \bibinfo{pages}{133604} (\bibinfo{year}{2016})}.

\bibitem{gonzalez_tudela2015}
\bibinfo{author}{Gonz\'alez-Tudela, A.}, \bibinfo{author}{Paulisch, V.},
  \bibinfo{author}{Chang, D.~E.}, \bibinfo{author}{Kimble, H.~J.} \&
  \bibinfo{author}{Cirac, J.~I.}
\newblock
\bibinfo{title}{Deterministic generation of arbitrary photonic states assisted
  by dissipation}.
  \href{http://dx.doi.org/10.1103/PhysRevLett.115.163603}{\newblock
  \emph{\bibinfo{journal}{Phys. Rev. Lett.}} \textbf{\bibinfo{volume}{115}},
  \bibinfo{pages}{163603} (\bibinfo{year}{2015})}.

\bibitem{Daley14}
\bibinfo{author}{Daley, A.~J.}
\newblock
\bibinfo{title}{Quantum trajectories and open many-body quantum systems}.
  \href{http://dx.doi.org/10.1080/00018732.2014.933502}{\newblock
  \emph{\bibinfo{journal}{Advances in Physics}} \textbf{\bibinfo{volume}{63}},
  \bibinfo{pages}{77--149} (\bibinfo{year}{2014})}.

\bibitem{Tsoi08}
\bibinfo{author}{Tsoi, T.~S.} \& \bibinfo{author}{Law, C.~K.}
\newblock
\bibinfo{title}{Quantum interference effects of a single photon interacting
  with an atomic chain inside a one-dimensional waveguide}.
  \href{http://dx.doi.org/10.1103/PhysRevA.78.063832}{\newblock
  \emph{\bibinfo{journal}{Phys. Rev. A}} \textbf{\bibinfo{volume}{78}},
  \bibinfo{pages}{063832} (\bibinfo{year}{2008})}.

\bibitem{Hafezi12}
\bibinfo{author}{Hafezi, M.}, \bibinfo{author}{Chang, D.~E.},
  \bibinfo{author}{Gritsev, V.}, \bibinfo{author}{Demler, E.} \&
  \bibinfo{author}{Lukin, M.~D.}
\newblock
\bibinfo{title}{Quantum transport of strongly interacting photons in a
  one-dimensional nonlinear waveguide}.
  \href{http://dx.doi.org/10.1103/PhysRevA.85.013822}{\newblock
  \emph{\bibinfo{journal}{Phys. Rev. A}} \textbf{\bibinfo{volume}{85}},
  \bibinfo{pages}{013822} (\bibinfo{year}{2012})}.

\bibitem{Znidaric15}
\bibinfo{author}{\ifmmode \check{Z}\else
  \v{Z}\fi{}nidari\ifmmode~\check{c}\else \v{c}\fi{}, M.}
\newblock
\bibinfo{title}{Relaxation times of dissipative many-body quantum systems}.
  \href{http://dx.doi.org/10.1103/PhysRevE.92.042143}{\newblock
  \emph{\bibinfo{journal}{Phys. Rev. E}} \textbf{\bibinfo{volume}{92}},
  \bibinfo{pages}{042143} (\bibinfo{year}{2015})}.

\bibitem{barends14}
\bibinfo{author}{Barends, R.} \emph{et~al.}
\newblock
\bibinfo{title}{Superconducting quantum circuits at the surface code threshold
  for fault tolerance}. \href{http://dx.doi.org/10.1038/nature13171}{\newblock
  \emph{\bibinfo{journal}{Nature}} \textbf{\bibinfo{volume}{508}},
  \bibinfo{pages}{500--503} (\bibinfo{year}{2014})}.

\bibitem{jeffrey14}
\bibinfo{author}{Jeffrey, E.} \emph{et~al.}
\newblock
\bibinfo{title}{Fast accurate state measurement with superconducting qubits}.
  \href{http://dx.doi.org/10.1103/PhysRevLett.112.190504}{\newblock
  \emph{\bibinfo{journal}{Phys. Rev. Lett.}} \textbf{\bibinfo{volume}{112}},
  \bibinfo{pages}{190504} (\bibinfo{year}{2014})}.

\bibitem{reagor18}
\bibinfo{author}{Reagor, M.} \emph{et~al.}
\newblock
\bibinfo{title}{Demonstration of universal parametric entangling gates on a
  multi-qubit lattice}.
  \href{http://dx.doi.org/10.1126/sciadv.aao3603}{\newblock
  \emph{\bibinfo{journal}{Science Advances}} \textbf{\bibinfo{volume}{4}},
  \bibinfo{pages}{eaao3603} (\bibinfo{year}{2018})}.

\bibitem{mckay17}
\bibinfo{author}{McKay, D.~C.}, \bibinfo{author}{Sheldon, S.},
  \bibinfo{author}{Smolin, J.~A.}, \bibinfo{author}{Chow, J.~M.},  \&
  \bibinfo{author}{Gambetta, J.~M.}
\newblock
\bibinfo{title}{Three qubit randomized benchmarking}. \newblock
  \emph{\bibinfo{journal}{arXiv}} \textbf{\bibinfo{volume}{1712.06550}}
  (\bibinfo{year}{2017}).

\bibitem{Menzel10}
\bibinfo{author}{Menzel, E.~P.} \emph{et~al.}
\newblock
\bibinfo{title}{Dual-path state reconstruction scheme for propagating quantum
  microwaves and detector noise tomography}.
  \href{http://dx.doi.org/10.1103/PhysRevLett.105.100401}{\newblock
  \emph{\bibinfo{journal}{Phys. Rev. Lett.}} \textbf{\bibinfo{volume}{105}},
  \bibinfo{pages}{100401} (\bibinfo{year}{2010})}.

\bibitem{wallraff2011}
\bibinfo{author}{Lang, C.} \emph{et~al.}
\newblock
\bibinfo{title}{Observation of resonant photon blockade at microwave
  frequencies using correlation function measurements}.
  \href{http://dx.doi.org/10.1103/PhysRevLett.106.243601}{\newblock
  \emph{\bibinfo{journal}{Phys. Rev. Lett.}} \textbf{\bibinfo{volume}{106}},
  \bibinfo{pages}{243601} (\bibinfo{year}{2011})}.

\bibitem{Hoi12}
\bibinfo{author}{Hoi, I.-C.}, \bibinfo{author}{Palomaki, T.},
  \bibinfo{author}{Lindkvist, J.}, \bibinfo{author}{Johansson, G.},
  \bibinfo{author}{Delsing, P.} \& \bibinfo{author}{Wilson, C.~M.}
\newblock
\bibinfo{title}{Generation of nonclassical microwave states using an artificial
  atom in {1D} open space}.
  \href{http://dx.doi.org/10.1103/PhysRevLett.108.263601}{\newblock
  \emph{\bibinfo{journal}{Phys. Rev. Lett.}} \textbf{\bibinfo{volume}{108}},
  \bibinfo{pages}{263601} (\bibinfo{year}{2012})}.

\bibitem{houck2017}
\bibinfo{author}{Fitzpatrick, M.}, \bibinfo{author}{Sundaresan, N.~M.},
  \bibinfo{author}{Li, A. C.~Y.}, \bibinfo{author}{Koch, J.} \&
  \bibinfo{author}{Houck, A.~A.}
\newblock
\bibinfo{title}{Observation of a dissipative phase transition in a
  one-dimensional circuit {QED} lattice}.
  \href{http://dx.doi.org/10.1103/PhysRevX.7.011016}{\newblock
  \emph{\bibinfo{journal}{Phys. Rev. X}} \textbf{\bibinfo{volume}{7}},
  \bibinfo{pages}{011016} (\bibinfo{year}{2017})}.

\bibitem{blais2010}
\bibinfo{author}{da~Silva, M.~P.}, \bibinfo{author}{Bozyigit, D.},
  \bibinfo{author}{Wallraff, A.} \& \bibinfo{author}{Blais, A.}
\newblock
\bibinfo{title}{Schemes for the observation of photon correlation functions in
  circuit {QED} with linear detectors}.
  \href{http://dx.doi.org/10.1103/PhysRevA.82.043804}{\newblock
  \emph{\bibinfo{journal}{Phys. Rev. A}} \textbf{\bibinfo{volume}{82}},
  \bibinfo{pages}{043804} (\bibinfo{year}{2010})}.

\bibitem{DiCandia14}
\bibinfo{author}{Candia, R.~D.}, \bibinfo{author}{Menzel, E.~P.},
  \bibinfo{author}{Zhong, L.}, \bibinfo{author}{Deppe, F.},
  \bibinfo{author}{Marx, A.}, \bibinfo{author}{Gross, R.} \&
  \bibinfo{author}{Solano, E.}
\newblock
\bibinfo{title}{Dual-path methods for propagating quantum microwaves}.
  \href{http://dx.doi.org/10.1088/1367-2630/16/1/015001}{\newblock
  \emph{\bibinfo{journal}{New J. Phys.}} \textbf{\bibinfo{volume}{16}},
  \bibinfo{pages}{015001} (\bibinfo{year}{2014})}.

\bibitem{Ramos17}
\bibinfo{author}{Ramos, T.} \& \bibinfo{author}{Garc\'{\i}a-Ripoll, J.~J.}
\newblock
\bibinfo{title}{Multiphoton scattering tomography with coherent states}.
  \href{http://dx.doi.org/10.1103/PhysRevLett.119.153601}{\newblock
  \emph{\bibinfo{journal}{Phys. Rev. Lett.}} \textbf{\bibinfo{volume}{119}},
  \bibinfo{pages}{153601} (\bibinfo{year}{2017})}.

\bibitem{Bohnet16}
\bibinfo{author}{Bohnet, J.~G.}, \bibinfo{author}{Sawyer, B.~C.},
  \bibinfo{author}{Britton, J.~W.}, \bibinfo{author}{Wall, M.~L.},
  \bibinfo{author}{Rey, A.~M.}, \bibinfo{author}{Foss-Feig, M.} \&
  \bibinfo{author}{Bollinger, J.~J.}
\newblock
\bibinfo{title}{Quantum spin dynamics and entanglement generation with hundreds
  of trapped ions}. \href{http://dx.doi.org/10.1126/science.aad9958}{\newblock
  \emph{\bibinfo{journal}{Science}} \textbf{\bibinfo{volume}{352}},
  \bibinfo{pages}{1297--1301} (\bibinfo{year}{2016})}.

\bibitem{luschen17}
\bibinfo{author}{L\"uschen, H.~P.} \emph{et~al.}
\newblock
\bibinfo{title}{Signatures of many-body localization in a controlled open
  quantum system}. \href{http://dx.doi.org/10.1103/PhysRevX.7.011034}{\newblock
   \emph{\bibinfo{journal}{Phys. Rev. X}} \textbf{\bibinfo{volume}{7}},
  \bibinfo{pages}{011034} (\bibinfo{year}{2017})}.

\bibitem{sieberer16}
\bibinfo{author}{Sieberer, L.~M.}, \bibinfo{author}{Buchhold, M.} \&
  \bibinfo{author}{Diehl, S.}
\newblock
\bibinfo{title}{Keldysh field theory for driven open quantum systems}.
  \href{http://dx.doi.org/10.1088/0034-4885/79/9/096001}{\newblock
  \emph{\bibinfo{journal}{Rep. Prog. Phys.}} \textbf{\bibinfo{volume}{79}},
  \bibinfo{pages}{096001} (\bibinfo{year}{2016})}.

\bibitem{fossfeig17}
\bibinfo{author}{Foss-Feig, M.}, \bibinfo{author}{Young, J.~T.},
  \bibinfo{author}{Albert, V.~V.}, \bibinfo{author}{Gorshkov, A.~V.} \&
  \bibinfo{author}{Maghrebi, M.~F.}
\newblock
\bibinfo{title}{Solvable family of driven-dissipative many-body systems}.
  \href{http://dx.doi.org/10.1103/PhysRevLett.119.190402}{\newblock
  \emph{\bibinfo{journal}{Phys. Rev. Lett.}} \textbf{\bibinfo{volume}{119}},
  \bibinfo{pages}{190402} (\bibinfo{year}{2017})}.

\bibitem{henriet18}
\bibinfo{author}{{Henriet}, L.}, \bibinfo{author}{{Douglas}, J.~S.},
  \bibinfo{author}{{Chang}, D.~E.} \& \bibinfo{author}{{Albrecht}, A.}
\newblock
\bibinfo{title}{{Critical open-system dynamics in a one-dimensional optical
  lattice clock}}.
  \href{http://dx.doi.org/10.1103/PhysRevA.99.023802}{\newblock
  \emph{\bibinfo{journal}{Phys. Rev. A}} \textbf{\bibinfo{volume}{99}},
  \bibinfo{pages}{023802} (\bibinfo{year}{2019})}.

\end{thebibliography}
\end{document}